\PassOptionsToPackage{numbers,sort&compress}{natbib}
\documentclass[preprintnumbers,amsmath,amssymb,prd, notitlepage,nofootinbib,twocolumn, superscriptaddress]{revtex4}
\usepackage{amsfonts,amssymb,amsmath}
\usepackage{gensymb}
\usepackage{color}
\usepackage{graphicx}
\usepackage{multirow}
\usepackage[utf8]{inputenc}
\usepackage{aas_macros}
\usepackage{hyperref}

\graphicspath{{./Images/}}
\usepackage[normalem]{ulem}

\newcommand{\greaterthanapprox}{\mathrel{\vcenter{
  \offinterlineskip\halign{\hfil$##$\cr
    >\cr\noalign{\kern2pt}\sim\cr\noalign{\kern-2pt}}}}}
    
    \newcommand{\lessthanapprox}{\mathrel{\vcenter{
  \offinterlineskip\halign{\hfil$##$\cr
    <\cr\noalign{\kern2pt}\sim\cr\noalign{\kern-2pt}}}}}
    
\newcommand{\lb}{\left(}
\newcommand{\rb}{\right)}

\newcommand{\Planck}{{\it Planck}~}

\newcommand{\be}{\begin{equation}}        
\newcommand{\ee}{\end{equation}}

\newcommand{\Neff}{N_\mathrm{eff}}

\newcommand{\lensed}{\mathrm{lensed}}

\begin{document}

\title{Baryonic feedback biases on fundamental physics from lensed CMB power spectra}

\author{Fiona McCarthy}
\email{fmccarthy@perimeterinstitute.ca}

\affiliation{Perimeter Institute for Theoretical Physics, Waterloo, Ontario, N2L 2Y5, Canada}
\affiliation{Department of Physics and Astronomy, University of Waterloo, Waterloo, Ontario, N2L 3G1, Canada }

\author{J.~Colin Hill}
\affiliation{Department of Physics, Columbia University, 538 West 120th Street, New York, NY, USA 10027}
\affiliation{Center for Computational Astrophysics, Flatiron Institute, New York, NY, USA 10003}

\author{Mathew S.~Madhavacheril}
\affiliation{Perimeter Institute for Theoretical Physics, Waterloo, Ontario, N2L 2Y5, Canada}

\date{\today}

\begin{abstract}
 
 Upcoming measurements of the small-scale primary cosmic microwave background (CMB) temperature and polarization power spectra ($TT$/$TE$/$EE$) are anticipated to yield transformative constraints on new physics, including the effective number of relativistic species in the early universe ($N_{\rm eff}$).  However, at multipoles $\ell \gtrsim 3000$, the primary CMB power spectra receive significant contributions from gravitational lensing.  While these modes still carry primordial information, their theoretical modeling requires knowledge of the CMB lensing convergence power spectrum, $C_L^{\kappa\kappa}$, including on small scales where it is affected by nonlinear gravitational evolution and baryonic feedback processes.  Thus, the high-$\ell$ primary CMB is sensitive to these late-time, nonlinear effects.  Here, we show that inaccuracies in the modeling of $C_L^{\kappa\kappa}$ can yield surprisingly large biases on cosmological parameters inferred from the primary CMB power spectra measured by the upcoming Simons Observatory and CMB-S4 experiments.  For CMB-S4, the biases can be as large as $1.6\sigma$ on the Hubble constant $H_0$ in a fit to $\Lambda$CDM and $1.2\sigma$ on $N_{\rm eff}$ in a fit to $\Lambda$CDM+$N_{\rm eff}$.
 We show that these biases can be mitigated by explicitly discarding all $TT$ data at $\ell>3000$ or by marginalizing over parameters describing baryonic feedback processes, both at the cost of slightly larger error bars.  We also discuss an alternative, data-driven mitigation strategy based on delensing the CMB $T$ and $E$-mode maps.  Finally, we show that analyses of upcoming data will require Einstein-Boltzmann codes to be run with much higher numerical precision settings than is currently standard, so as to avoid similar --- or larger --- parameter biases due to inaccurate theoretical predictions.
 \end{abstract}
\maketitle

\section{Introduction}
Measurements of the cosmic microwave background (CMB) temperature and polarization anisotropy power spectra have revolutionized our understanding of cosmology in the past few decades~(e.g.,~\cite{2013ApJS..208...20B,Akrami:2018vks,2020JCAP...12..047A,2021arXiv210101684D}).  Upcoming CMB anisotropy measurements promise to build upon this success, with unprecedented sensitivity to signals of new physics in the early universe~\cite{Ade:2018sbj,CMBS4DSR}.  Key to this success is the robust theoretical foundation upon which CMB anisotropy power spectrum calculations rest.  In particular, the primary CMB fluctuations are described to very high accuracy by linear cosmological perturbation theory.  As first recognized long ago~(e.g.,~\cite{1996PhRvL..76.1007J,1996PhRvD..54.1332J,1997MNRAS.291L..33B,1997ApJ...488....1Z}), high-precision measurements combined with this robust theoretical foundation allow constraints on all of the fundamental cosmological parameters (in $\Lambda$CDM) to be inferred solely from the CMB.  Upcoming experiments will utilize this power to put leading constraints on many new-physics parameters, including the effective number of relativistic species ($N_{\rm eff}$), the sum of the neutrino masses ($M_{\nu}$), the running of the spectral index of primordial perturbations, models to resolve the $H_0$ tension, and many other scenarios.  All of these constraints rely on the precise modeling of the CMB power spectra within linear perturbation theory.

However, on small angular scales in the CMB, crucial assumptions in this picture begin to break down.  Gravitational lensing of CMB photons, which distorts their paths as they travel from the surface of last scatter to our telescopes, leads to subtle but non-negligible changes to the CMB power spectra~(for a comprehensive review of CMB lensing, see~\cite{Lewis:2006fu}).  In particular, gravitational lensing smooths the acoustic peaks and pushes anisotropy power into the high-multipole ``Silk damping tail'' of the CMB power spectra~\cite{1987A&A...184....1B,1996ApJ...463....1S,1998PhRvD..58b3003Z}.  Thus, due to lensing, the primary CMB is influenced by the properties of the matter density field at low redshifts, as captured in the CMB lensing potential field, which is a particular redshift-weighted projection of the density field along the line-of-sight (LOS).  The CMB lensing potential field is well-described by linear perturbation theory on angular scales greater than $\approx$ 10 arcmin~\cite{Lewis:2006fu}, but on smaller scales it is affected by nonlinear gravitational evolution and processes associated with baryons, such as feedback from active galactic nuclei (AGN)~\cite{2006ApJ...640L.119J,2008ApJ...672...19R,2011MNRAS.417.2020S,vanDaalen2011,2014PhRvD..90f3516N,2017MNRAS.465.2936M,2018MNRAS.475..676S,2018MNRAS.480.3962C,2019JCAP...03..020S,2020PhRvD.101f3534C,2020arXiv201106582M}.  Thus, one may wonder to what extent these highly nonlinear processes could affect the primary CMB itself through gravitational lensing, and whether these effects could influence cosmological parameter inference from upcoming high-resolution CMB experiments.

In this paper, we show that nonlinear and baryonic effects can indeed produce significant biases in the analysis of data from upcoming CMB experiments.  The small-scale (multipoles $\ell \gtrsim 3000$) CMB receives sufficiently large contributions from the small-scale ($L \gtrsim 2000$) CMB lensing potential field that these effects cannot be ignored.  We consider a range of models for the effects of nonlinear evolution and baryonic feedback on the small-scale CMB lensing power spectrum, and compute their effects on the primary CMB power spectra.  We then propagate these models through a Fisher analysis to forecast biases on cosmological parameters that would be inferred when assuming an incorrect (but currently standard) model.  For concision, we focus on the $\Lambda$CDM and $\Lambda$CDM+$\Neff$ models, where $\Neff$ is the effective number of relativistic species.  The latter is of particular interest, as constraints on $\Neff$ are strongly driven by measurements of the damping tail in the primary CMB power spectra, which is also the region most altered by the effects identified in this work.  However, similar biases for other parameters (e.g., the sum of the neutrino masses or the running of the spectral index) are also likely to exist, and should be considered (and mitigated) in upcoming CMB data analyses.

A brief summary of our results is as follows.  We show that constraints on $N_{\rm eff}$ from the upcoming Simons Observatory (SO)~\cite{Ade:2018sbj} and CMB-S4~\cite{CMBS4DSR} experiments could be biased by up to $0.4\sigma$ and $1.2\sigma$, respectively, due to the neglect of baryonic feedback in modeling of the primary CMB power spectra.  Similarly, constraints on the physical cold dark matter density, $\Omega_{\rm c} h^2$, could be biased by up to $1.0\sigma$ (SO) and $1.6\sigma$ (CMB-S4) in $\Lambda$CDM, or up to $0.8\sigma$ (SO) and $2.0\sigma$ (CMB-S4) in $\Lambda$CDM+$N_{\rm eff}$. The Hubble constant $H_0$, which is a derived parameter in the analysis of CMB data, could be biased by up to $1.0\sigma$ (SO) and $1.6\sigma$ (CMB-S4) in $\Lambda$CDM. In general, the bias on a given parameter depends on the model under consideration, as parameter degeneracies will change.
Table~\ref{tab:biases_SOS4} summarizes the biases on the cosmological parameters for SO and CMB-S4 in the $\Lambda$CDM and $\Lambda$CDM+$\Neff$ models.  As a byproduct of this analysis, we also investigate the effects of numerical precision errors in Einstein-Boltzmann codes (e.g., CAMB\footnote{\url{http://camb.info}}~\cite{2000ApJ...538..473L} or CLASS\footnote{\url{http://class-code.net/}}~\cite{Blas2011}) on the high-$\ell$ CMB, which can lead to similar (or even larger) parameter biases if increased accuracy settings are not adopted when running these codes.

We suggest multiple mitigation strategies to avoid these potentially significant baryonic feedback-induced biases.  The simplest approach is to explicitly discard all high-$\ell$ $TT$ power spectrum data.  At a fixed multipole in the damping tail, the small-scale $TT$ power spectrum is most affected by the lensing contributions described above (compared to $TE$ or $EE$), due to the larger gradient in the unlensed $T$ field as compared to $E$.  Moreover, due to the much larger amplitude of the $TT$ signal, CMB experiments measure more signal-dominated modes in the $TT$ damping tail than in $TE$ or $EE$ (even after accounting for foregrounds).  Thus, the biases that we compute are generally driven most strongly by $TT$.  Explicitly, we find that biases on all parameters investigated here can be kept to $\lesssim 0.3\sigma$ if all $TT$ data at $\ell \gtrsim 3000$ are discarded (see Figure~\ref{fig:lmaxcut_allsims} and Table \ref{tab:biases_SOS4_lmax3000}).  However, discarding the $TT$ data comes at the price of increased statistical error bars on cosmological parameters.  Fortunately, the increase is not dramatic: at most $\approx 21\%$ for $\Omega_c h^2$ and $\approx 13\%$ for $\Neff$.

\begin{table}[!t]
\begin{tabular}{|c|c|c|c|c|}\hline
&\multicolumn{2}{c|}{SO}&\multicolumn{2}{c|}{CMB-S4}\\\cline{2-5}
&$\Lambda$CDM&$\Lambda$CDM$+\Neff$&$\Lambda$CDM&$\Lambda$CDM$+\Neff$\\\hline\hline
$H_0$   &   0.96   &   0.15   &   1.6   &   0.035\\\hline
$\Omega_bh^2$   &   0.070   &   0.27   &   0.44   &   0.56\\\hline
$\Omega_ch^2$   &   1.0   &   0.96   &   1.6   &   2.0\\\hline
$	\tau$   &   0.37   &   0.42   &   0.28   &   0.42\\\hline
$A_s$   &   0.57   &   0.68   &   0.52   &   0.81\\\hline
$n_s$   &   0.36   &   0.16   &   0.48  &   0.69\\\hline
$N_{\mathrm{eff}}$   &      &   0.44   &      &   1.2\\\hline

\end{tabular}
\caption{Fractional biases (in units of the forecast $1\sigma$ statistical error bar) on each of the parameters in the various setups, if baryonic effects are ignored.  Note that the biases are different for the same parameters in the $\Lambda$CDM and $\Lambda$CDM+$\Neff$ models due to effects of the marginalization over $\Neff$. We assume a maximum multipole $\ell_{\rm max} = 5000$ here, with noise power spectra for SO and CMB-S4 shown in Figure~\ref{fig:lensed_unlensed_power}. The OWLS-AGN~\cite{vanDaalen2011,vanDaalen:2019pst} baryonic model is assumed here.}\label{tab:biases_SOS4}
\end{table}

Another mitigation approach is to explicitly parameterize the nonlinear/baryonic effects on the CMB lensing power spectrum, and subsequently marginalize over these parameters in the cosmological analysis of the primary CMB power spectra (e.g., as done in~\cite{2020arXiv201106582M} for the cosmological analysis of the CMB lensing power spectrum).  We perform this exercise in our Fisher calculations below.  We find that this approach can successfully mitigate the biases, but, like the approach suggested above, comes at the price of increased error bars on cosmological parameters.  However, we note that this can likely be improved by performing a joint analysis of the primary CMB power spectra with the reconstructed CMB lensing potential power spectrum itself.  This will require a precise treatment of the joint covariance between these probes~\cite{Peloton2017}.

Finally, the most data-driven approach would be to ``delens'' the temperature and polarization anisotropy maps using the measured CMB lensing potential, e.g., as reconstructed using quadratic estimators~\cite{2017PhRvD..95j3512S,2017JCAP...05..035C} or maximum-likelihood methods~\cite{2017PhRvD..96f3510C,2019PhRvD.100b3509M}, or as traced by external probes like the cosmic infrared background~\cite{2012JCAP...06..014S,Sherwin:2015baa,2016PhRvL.117o1102L,2017PhRvD..96l3511Y}.  If the delensing operation were 100\% efficient, then because delensing uses the observed (true) lensing potential field to undo the lensing effects, it is clear that all biases related to modeling of the late-time density field would be removed in the primary CMB power spectra (since no such modeling would be required).  Assessing the fidelity of this operation for realistic experimental configurations, which will yield less-than-perfect delensing efficiencies, will be a useful exercise in upcoming work.  In particular, the feasibility of delensing on such small scales has not yet been explored.  It is interesting to note that due to the effects identified here, in addition to statistical optimality arguments identified in earlier work~\cite{2017JCAP...12..005G}, delensing now appears to be an important operation not only for enabling precise constraints on the tensor-to-scalar ratio $r$~\cite{2012JCAP...06..014S,2015ApJ...807..166S,Ade:2018sbj,CMBS4DSR}, but also for enabling unbiased constraints on $N_{\rm eff}$ and other new-physics parameters in upcoming CMB experiments.

The remainder of this paper is organized as follows. In Section \ref{sec:theory} we discuss gravitational lensing of the CMB and how uncertainties in the lensing power spectrum can propagate to uncertainties in the \textit{lensed} power spectra. In Section \ref{sec:fisher} we introduce the Fisher formalism we use to forecast error bars and systematic biases. In Section \ref{sec:baryons} we discuss biases from the mismodeling of unknown baryonic effects, and present several ways to remove these biases. In Section \ref{sec:inaccuracy_bias} we discuss potential biases due to numerical precision errors in calculating the lensed power spectra, and quantify how this translates to biased parameter inferences.  We discuss our results and conclude in Section~\ref{sec:discussion}.

All of our power spectrum calculations are performed with CAMB \cite{2000ApJ...538..473L}. We assume a fiducial cosmology throughout of $\{H_0 = 67.32 \,\, \mathrm{km/s/Mpc}, \Omega_b h^2=0.022383,\Omega_c h^2=0.12011,n_s= 0.96605, A_s=2.1\times 10^{-9}, \tau = 0.0543, M_\nu=0.06 \,\, \mathrm{eV},\Neff = 3.046\}$, corresponding to the best-fit values of the six primary $\Lambda$CDM parameters found in Table 1 of \cite{2020A&A...641A...6P} along with the minimum allowed value of the neutrino mass $M_\nu$ and the standard value of the effective number of neutrino species $\Neff$ (which are the values assumed in \cite{2020A&A...641A...6P}).

\section{The lensed CMB power spectra}\label{sec:theory}

\begin{figure*}[t]
\includegraphics[width=0.49\textwidth]{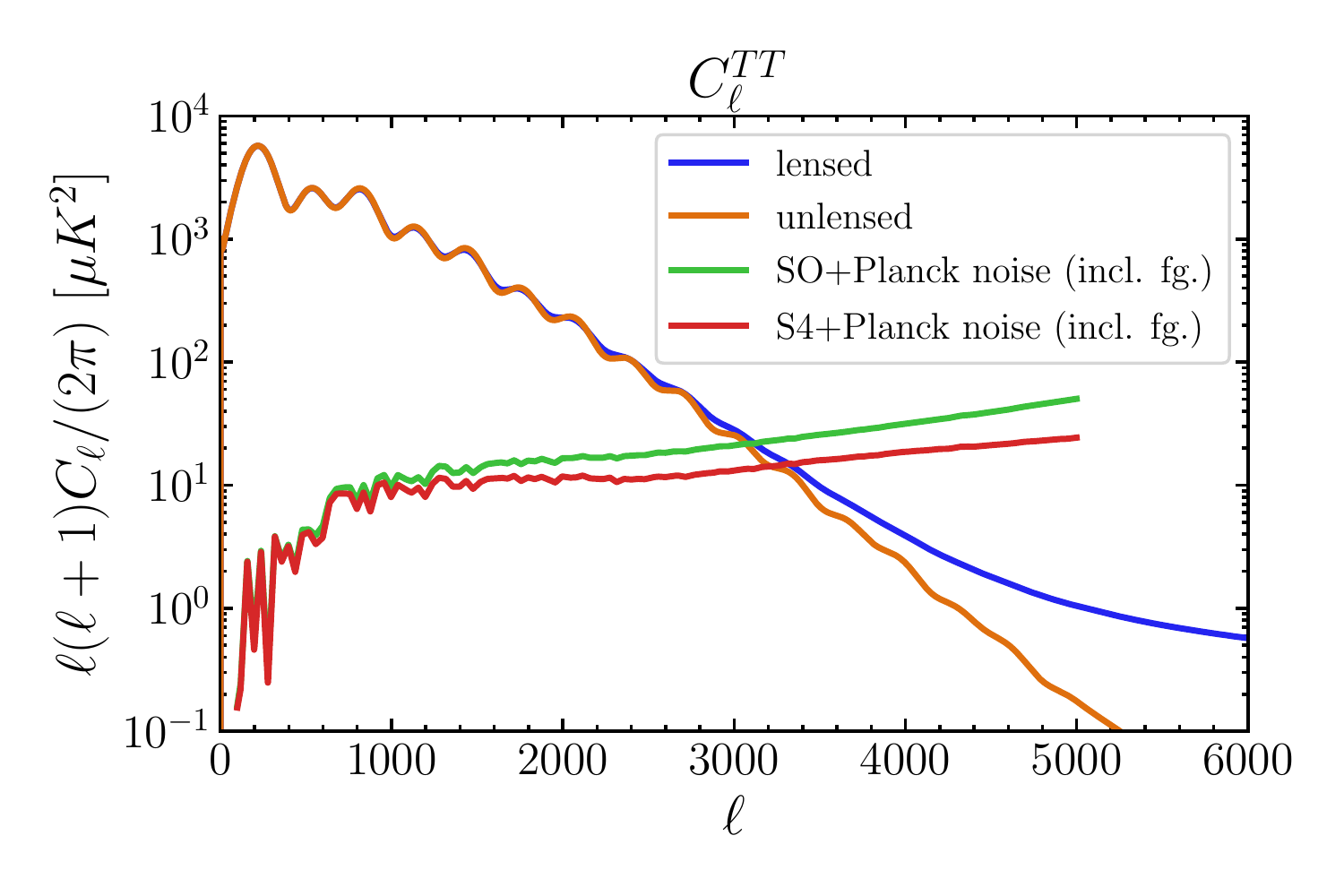}
\includegraphics[width=0.49\textwidth]{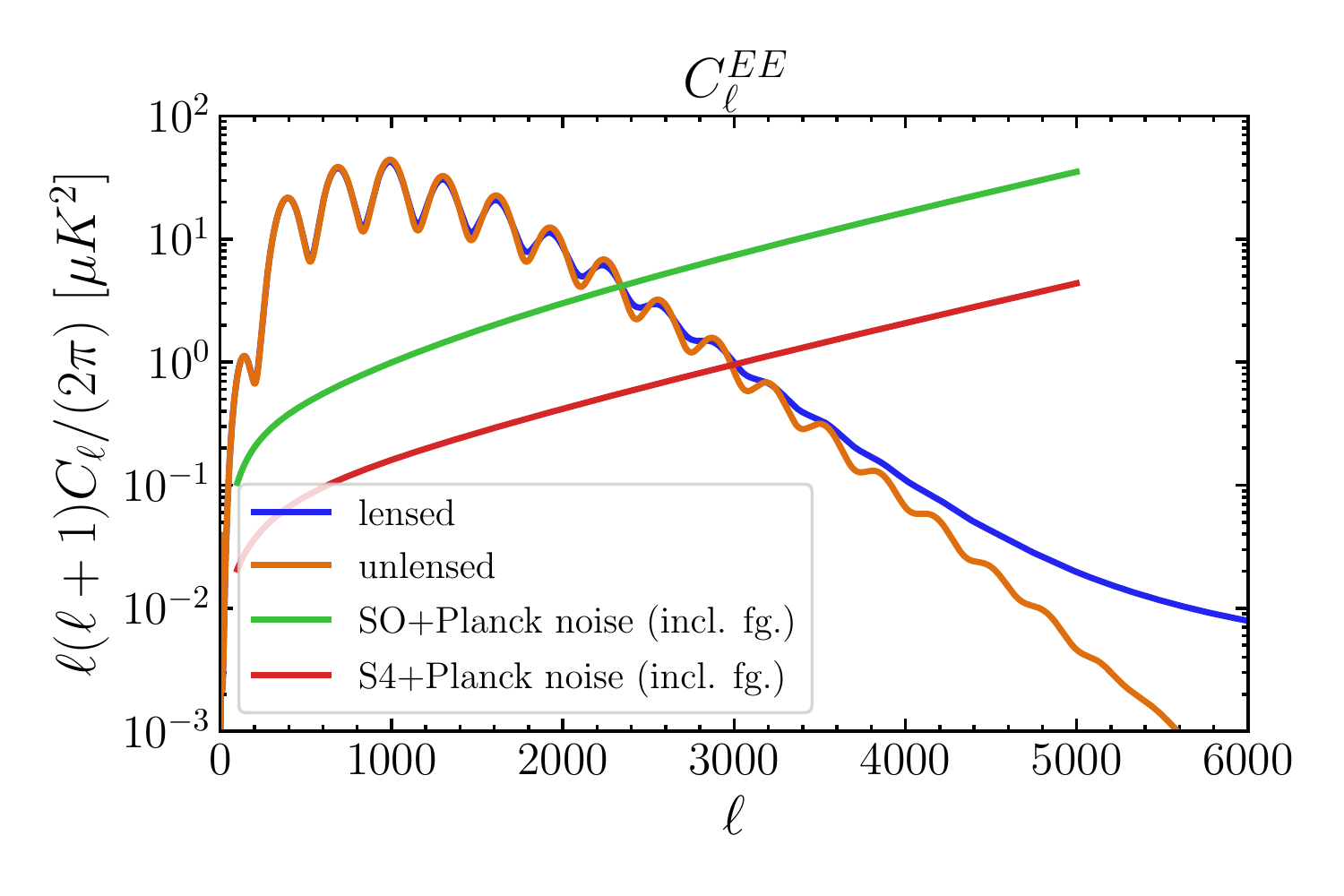}
\includegraphics[width=0.49\textwidth]{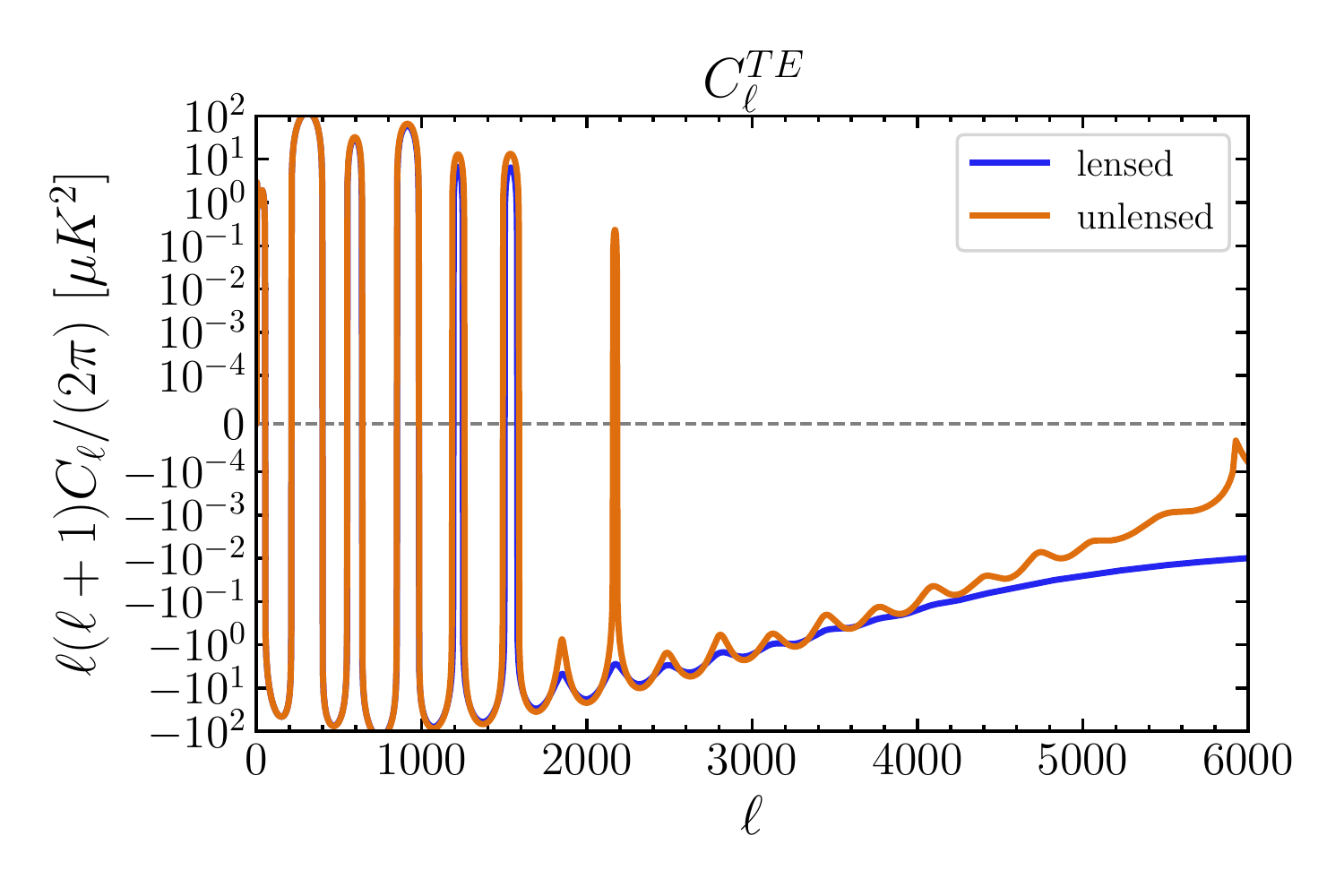}
\includegraphics[width=0.49\textwidth]{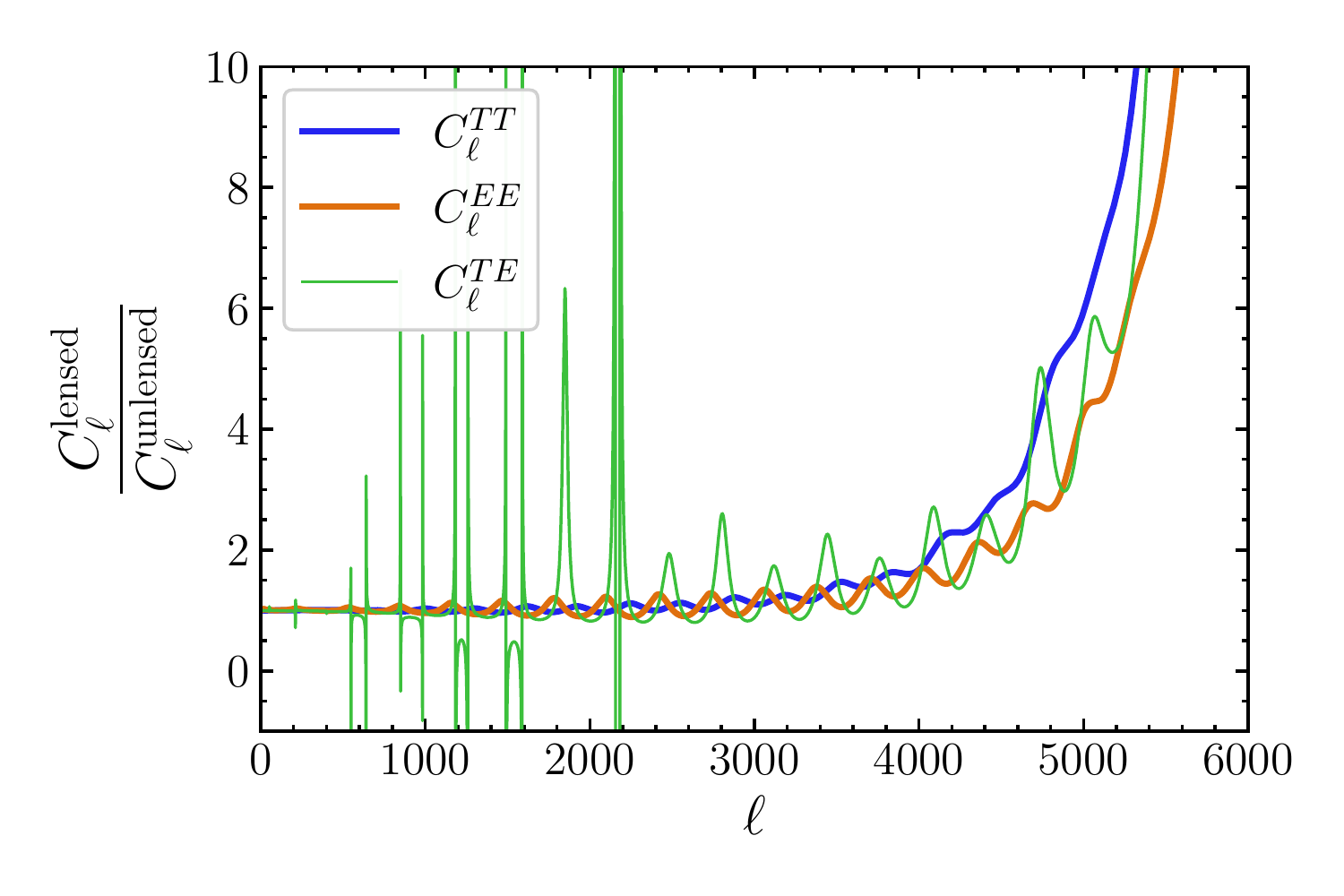}
\caption{The lensed (blue) and unlensed (orange) CMB $TT$ (\emph{top left}), $EE$ (\emph{top right}), and $TE$ (\emph{bottom left}) power spectra and their ratios ({\textit {bottom right})}. We see in the lensed spectra an increase of power on small scales, also clearly illustrated in the bottom right plot. We also see the smoothing effect of lensing, in the lowering of peak heights and the raising of trough heights. The post-component-separation noise power spectra (including residual foregrounds) expected from SO~\cite{Ade:2018sbj} and CMB-S4~\cite{CMBS4DSR}, both including \emph{Planck} data as well, are also indicated on the $C_\ell^{TT}$ and $C_\ell^{EE}$ plots (see Sec.~\ref{sec:fisher} for details).  It is clear that for precision cosmology with these experiments we will need to have an accurate calculation of the {\textit{lensed}} CMB power. }\label{fig:lensed_unlensed_power}
\end{figure*}

Significant cosmological analysis is performed with the two-point statistics of the observed CMB, for which we have well-understood theoretical predictions. In particular, we consider the power spectrum of the CMB intensity anisotropies, $C_\ell^{TT}$, and the power spectrum of the $E$-mode CMB polarization anisotropies, $C_\ell^{EE}$. As these probes are not fully independent (the CMB is roughly 10\% polarized), we also consider their cross-power spectrum, $C_\ell^{TE}$. The effect of gravitational lensing on these quantities is shown in Figure~\ref{fig:lensed_unlensed_power}.  We provide a brief summary of the relevant physics here, and refer the reader to Ref.~\cite{Lewis:2006fu} for full details. 

The lensed power spectra, $C_\ell^{\rm lensed}$, are functions of the unlensed power spectra $C_\ell^{\rm unlensed}$, and the lensing potential power spectrum, $C_L^{\phi\phi}$.  It is also common to consider, instead of $C_L^{\phi\phi}$, the lensing convergence power spectrum $C_L^{\kappa\kappa}$, which is related to $C_L^{\phi\phi}$ in harmonic space by
\be
C_L^{\kappa\kappa} = \frac{\lb L(L+1)\rb^2}{4}C_L^{\phi\phi}.
\ee
$C_L^{\kappa\kappa}$ is given in the Limber \cite{1953ApJ...117..134L} and Born approximations by
\be
C_L^{\kappa\kappa} = \int_0^{\chi_\mathrm{CMB}}   d\chi \left( \frac{W_{\mathrm {CMB}}^\kappa(\chi)}{\chi}\right)^2 P_m\lb k=\frac{L+1/2}{\chi},z\rb\label{clkappakappa},
\ee
where $P_m(k,z)$ is the matter power spectrum and $W_{\mathrm {CMB}}^\kappa(\chi)$ is the CMB lensing efficiency kernel
\be
W_{\mathrm{ CMB}}^\kappa(\chi) = \frac{3}{2}\Omega_m\lb \frac{H_0}{c}\rb^2\frac{\chi}{a(\chi)}\frac{\chi_{\rm CMB}-\chi}{\chi_{ \rm CMB}}\label{efficiency_CMB}
\ee
with $\chi_{\rm CMB}$ the comoving distance to the surface of last scattering.

The lensing potential affects the observed temperature anisotropies. In particular, when we look in a direction $\hat n$, we do not observe the temperature emitted at $\hat n$ but the temperature that has been lensed into that direction, which was in fact sourced in the direction $\hat n + \vec \alpha$, where the total deflection angle induced by lensing $\vec \alpha$ is given by the gradient (in the plane of the sky) of the lensing potential $\phi$. Section 4 of Ref.~\cite{Lewis:2006fu} provides a thorough review of the calculation of the $C_\ell^{\mathrm{lensed}}$ from $C_\ell^{\mathrm{unlensed}}$ and $C_\ell^{\phi\phi}$, and we refer the interested reader to Section 4.2 of that paper for details of the exact calculation. Note, however, that to first order in $C_\ell^{\phi\phi}$ ($\sim$ to second order in $\vec \alpha$) the lensed temperature power spectrum can be expressed as a convolution between the unlensed temperature power spectrum and the lensing potential power spectrum
\begin{align}
C_\ell^{TT}{}^{\mathrm{lensed}} \approx \lb 1-\ell^2 R^\phi\rb C_\ell^{TT}{}^{\mathrm{unlensed}}\nonumber\\
+\int \frac{d^2 \vec \ell^\prime}{\lb2\pi\rb^2}C^{\phi\phi}_{\left |\vec \ell - \vec \ell^\prime\right |}C_{\ell^\prime}^{TT}{}^{\mathrm{unlensed}}
\label{eq.ClTTlensed_highell}
\end{align}
where $R^\phi$ is given by
\be
R^\phi \equiv \frac{1}{4\pi}\int\frac{d\ell}{\ell}\ell^4 C_\ell^{\phi\phi}.
\ee
In the small-scale limit, while the expansion in small $C_\ell^{\phi\phi}$ is not accurate, the fact that the unlensed power is so small and can be described by a single gradient term also allows for an approximation:
\be
C_\ell^{TT}{}^{\mathrm{lensed}} \approx \ell^2 C_\ell^{\phi\phi}R^{\Theta}
\ee
where
\be
R^{\Theta} = \frac{1}{4\pi}\int \frac{d\ell}{\ell}\ell^4 C_\ell^{TT}{}^{\mathrm{unlensed}} \,.
\ee
For details of the above approximations we refer the reader to Section 4.1 of \cite{Lewis:2006fu} and references therein; the calculation of the exact lensed $TT$ spectrum by means of the correlation function, which is the method used in CAMB and CLASS, is also discussed in Section 4.2 therein. Similarly, the lensing of the $EE$ spectrum is discussed in Section 5.3 of \cite{Lewis:2006fu}. 
There are a number of important aspects to consider in the accurate calculation of $C_L^{\kappa\kappa}$ (and thus the accurate calculation of the lensed $C_\ell^{TT}$, $C_\ell^{TE}$, and $C_\ell^{EE}$).  Upcoming CMB surveys, including SO and CMB-S4, will be sufficiently sensitive that seemingly small effects need to be taken into account (our modeling of the noise for these surveys is discussed in the next section).  First, inaccuracies in the modeling of $P_m(k,z)$ due to gravitational non-linearities and baryonic feedback effects will become sufficiently important as to affect not only the interpretation of the reconstructed lensing power spectrum~\cite{2020PhRvD.101f3534C}, but also the lensed primary CMB power spectra themselves.  Second, the inferences we will make with upcoming CMB surveys will be sensitive to the numerical precision parameters used in the calculation; in particular the default precision settings of CAMB will be insufficient for SO and CMB-S4.  We will quantify these statements in the following sections. 

The linear matter power spectrum is computed with cosmological perturbation theory, with non-linearities incorporated through, e.g., a halo-model-based fitting function such as Halofit~\cite{Takahashi2012,Smith2003} or HMCode~\cite{2015MNRAS.454.1958M,2021MNRAS.502.1401M}.  HMCode further includes free parameters intended to capture the effects of complex baryonic phenomena on the matter power spectrum, including gas cooling and AGN and supernova feedback.  These baryonic effects significantly alter the clustering of matter on $\lesssim 10 \, \rm{Mpc}$ scales.  We do not currently have a first-principles calculation of such effects, and as such they are sources of systematic error in the modeling of non-linear power spectra.  To gain some understanding of the effects of baryonic interactions on matter clustering, we can perform cosmological hydrodynamics simulations~(e.g.,~\cite{2017MNRAS.465.2936M,2018MNRAS.475..676S,2018MNRAS.480.3962C}); indeed, hydrodynamical simulations are used to construct the parametric HMCode model.  However, one should keep in mind that the true non-linear power spectrum in our universe could (and likely does) differ at some level from these models (see, e.g., \cite{2009.05557,2009.05558}).  As we will show, accounting for this uncertainty will be important in upcoming CMB experiments focused on the small-scale \emph{primary} CMB power spectra.

\section{Inference of cosmological parameters: statistical and systematic errors}\label{sec:fisher}

\subsection{The Fisher matrix formalism}

The Fisher matrix formalism is widely used to calculate the uncertainties expected from statistical error alone on the analysis of a given parameter, assuming a (theoretical) calculation of the covariance of the data expected, including noise contributions.  We summarize this approach briefly here.

We take as a data vector the (lensed) CMB power spectra:
\be 
C_\ell = \left\{C_\ell^{\lensed}{}^{TT},C_\ell^{\lensed}{}^{EE},C_\ell^{\lensed}{}^{TE}\right\}.
\ee
We calculate the theoretical $C_\ell$ with CAMB. The covariances between the different $C_\ell$'s are given by
\begin{widetext}
\begin{align}
\mathbb{C}\left(\hat{C}_{\ell}^{\alpha \beta}, \hat{C}_{\ell}^{\gamma \delta}\right)= \frac{1}{(2 \ell+1) f_{\mathrm{sky}}}\bigg{[}\left(C_{\ell}^{\alpha \gamma}
+ N_{\ell}^{\alpha \gamma}\right)\left(C_{\ell}^{\beta \delta}+N_{\ell}^{\beta \delta}\right)
+\left(C_{\ell}^{\alpha \delta}+N_{\ell}^{\alpha \delta}\right)\left(C_{\ell}^{\beta \gamma}+N_{\ell}^{\beta \gamma}\right)\bigg{]},\label{cov_bandpower}
\end{align}
\end{widetext}
where $N_\ell^{XY}$ is the noise on the measurement of $C_\ell^{XY}$, including contributions from the instrument, atmosphere, and residual foregrounds after component separation. We assume the noise on the polarization and intensity measurements to be uncorrelated, i.e., $N_\ell^{TE} = 0$.  Finally, in Eq.~\ref{cov_bandpower} $f_{\rm sky}$ is the sky fraction covered by the survey; we take $f_{\rm sky}=0.4$ for SO and $f_{\rm sky}=0.45$ for CMB-S4. 

We include post-component-separation noise power spectra $N_\ell^{TT}$ and $N_\ell^{EE}$, as computed for either SO (using the ``Goal'' noise levels)\footnote{\url{https://simonsobservatory.org/assets/supplements/20180822_SO_Noise_Public.tgz}}~\cite{Ade:2018sbj} or CMB-S4\footnote{\url{https://cmb-s4.uchicago.edu/wiki/index.php/Survey_Performance_Expectations}}~\cite{CMBS4DSR}, both in combination with \textit{Planck} data.  The noise power spectra include contributions from instrumental and atmospheric noise, as well as residual foregrounds after multi-frequency internal linear combination (ILC) foreground cleaning has been applied in the harmonic domain~(e.g.,~\cite{2003PhRvD..68l3523T,2004ApJ...612..633E}).  The foregrounds include models for Galactic dust and synchrotron in both temperature and polarization, as well as Galactic free-free, Galactic spinning dust, extragalactic radio and infrared point sources, the thermal and kinematic Sunyaev-Zel'dovich effects, and the cosmic infrared background in temperature, with realistic correlations amongst the constituent Galactic and extragalactic components (full details can be found in~\cite{Ade:2018sbj,CMBS4DSR}).  While the modeling of these components is not perfectly known, this uncertainty will only affect the post-component-separation noise power spectra in Fig.~\ref{fig:lensed_unlensed_power} at the $\sim 10$\% level; the biases computed in this paper will thus be expected to differ slightly in practice compared to our forecasts, but not dramatically so.  However, we note that in a fully realistic analysis of multi-frequency power spectrum data, the contributions from various foregrounds would be parameterized and explicitly marginalized over in the likelihood (e.g.,~\cite{2013JCAP...07..025D,2020A&A...641A...5P}).  If the foregrounds are sufficiently orthogonal to the primary CMB (as is the case with non-blackbody foregrounds probed in multiple frequency channels) and the model has sufficient flexibility, then the foregrounds will not bias cosmological parameter estimation, and their effect is primarily to contribute to the noise power captured in our post-ILC noise curves.\footnote{This statement was explicitly verified for SO forecasts by comparing the post-ILC effective noise curve approach to a full, parametric likelihood calculation in Sec.~4.1.2 of Ref.~\cite{Ade:2018sbj}.}   Non-Gaussian contributions of the foregrounds to the post-ILC noise covariance matrix, as well as the question of whether currently used foreground models are sufficiently flexible so as to not bias cosmological parameters are outside the scope of this work.  Similarly, in this work, we do not include contributions from non-Gaussian covariance due to lensing and super-sample variance \cite{Peloton2017,GreenDelensing,2019PhRvD..99b3506M}, which would especially be of importance in a mitigation strategy involving joint analysis with the CMB lensing four-point function (see Sec.~\ref{sec:baryonmarg}).

The Fisher matrix for the parameter vector $ \Pi^i$ can be calculated from the covariance $\mathbb{C}_\ell$ and derivatives of the data vector $C_\ell$ with respect to $\Pi^i$:\footnote{Here we have assumed that the covariance matrix $\mathbb{C}$ does not depend on the parameters, i.e., it is computed at a fixed cosmology.}
\be
F_{ij}=\sum_\ell \frac{\partial C_\ell^T}{\partial \Pi^i }\mathbb{C}_\ell^{-1} \frac{\partial C_\ell}{\partial \Pi^j }.\label{fisher_matrix}
\ee
Within this formalism, the forecast statistical error on the parameter $i$, marginalized over the other parameters in $\Pi$, is
\be
\sigma(\Pi^i) = \sqrt{\lb F^{-1}\rb_{ii}}.\label{sigma}
\ee
Note that Eq.~\eqref{sigma} represents a \textit{lower bound} on the true error bars, with the true error bar approaching $\sqrt{\lb F^{-1}\rb_{ii}}$ in the case of Gaussian covariances.

The standard Fisher formalism above can be extended to consider errors that are not statistical, but instead are caused by a systematic miscalculation of the theoretical signal, e.g., an incorrect theory model. If we perform data analysis with incorrect theoretical power spectra --- let us call this $C_\ell^{\mathrm {fiducial}}$ --- the dependence of $C_\ell^{\mathrm{ fiducial}}$ on the parameters $\Pi^i$ will be different to those of the true theory ($C_{\ell}^{\rm true}$), and we will get a biased inference of $\Pi^i$.  The size of the bias is given by the bias vector (see, e.g., \cite{2005APh....23..369H,2008MNRAS.391..228A})
\be
B(\Pi^i) = F_{ij}^{-1}\sum_\ell \frac{\partial C_\ell^{{\rm fiducial},T}}{\partial \Pi^j }\mathbb{C}_\ell^{-1} \Delta C_\ell,\label{bias}
\ee
where 
\be
\Delta C_\ell \equiv C_\ell^{\mathrm{true}}-C_\ell^{\mathrm{fiducial}}.
\ee
Eq.~\eqref{bias} thus allows us to compute \textit{systematic} errors in the Fisher formalism, arising due to differences between the correct and assumed theoretical power spectra. 

\subsection{Constraints from upcoming surveys}

\begin{figure}[!t]
\includegraphics[width=\columnwidth]{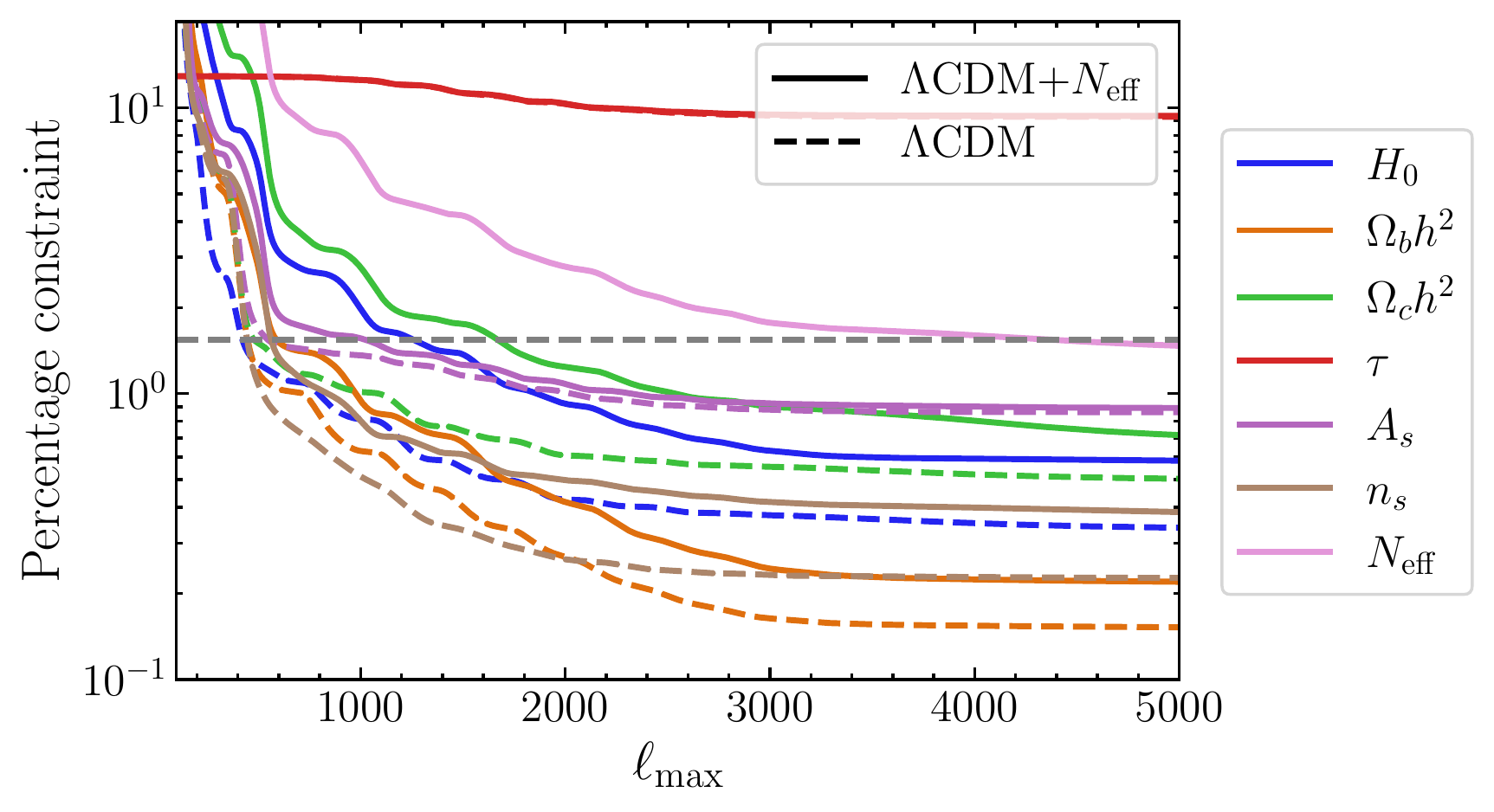}
\caption{The forecast CMB-S4 constraints on the parameters of the $\Lambda$CDM model (dashed) and of $\Lambda$CDM$+\Neff$ (solid), plotted against the maximum multipole included in the analysis. The constraints are shown as the fractional statistical uncertainty on each parameter, in units of percent.}\label{fig:constraints_LCDM}
\end{figure}

We can use the Fisher formalism to predict the statistical error bars on cosmological parameters for upcoming experiments, e.g., SO and CMB-S4. In particular, we show in Figure \ref{fig:constraints_LCDM} the predicted CMB-S4 constraints on the six parameters of the $\Lambda$CDM model, and on an extension of this model where the effective number of relativistic species $\Neff$ is allowed to vary. We sum over all multipoles from $\ell_{\mathrm{min}}=100$ to $\ell_{\mathrm{max}}$, the quantity labeled on the $x$-axis.  We also include a Gaussian prior on the parameter $\tau$ of $0.0543 \pm 0.007$, which is constrained to this level by the large-scale \Planck $EE$ data~\cite{2020A&A...641A...6P}.\footnote{The SO and CMB-S4 large aperture telescopes are not expected to measure the largest-scale modes on the sky due to atmospheric $1/f$ noise~\cite{Ade:2018sbj,CMBS4DSR}, which necessitates the use of our prior on $\tau$ here.}  From the figure, it is clear that CMB-S4 can constrain all parameters in the base $\Lambda$CDM model to sub-percent precision (except $\tau$), and $N_{\rm eff}$ to near-percent precision, solely using the primary CMB power spectra.  As expected, the constraints begin to saturate once $\ell_{\rm max}$ is greater than the multipole at which CMB-S4 is cosmic-variance-limited (roughly $\ell \approx 3500$ in $TT$ and $\ell \approx 3000$ in $EE$, as shown in Fig.~\ref{fig:lensed_unlensed_power}).  For experiments with even lower noise levels than CMB-S4, these constraints would continue to improve with increasing $\ell_{\rm max}$~(e.g.,~\cite{CMBHD,PICO}).  While we only show the $\ell_{\mathrm{max}}$-dependence of the CMB-S4 statistical forecast here for brevity, we note that SO will yield very precise constraints as well (prior to the start of the CMB-S4 survey), e.g., with a forecast error bar on $N_{\rm eff}$ of roughly 2\%~\cite{Ade:2018sbj}; the values of the constraints that we calculate are listed in Tables 
\ref{tab:LCDM_forecast_full} and \ref{tab:LCDMNeff_forecast_full} in Appendix~\ref{app:biases}.  Thus, it is well-motivated to consider both experiments in our analysis.

\section{Physical systematics: mismodeling of baryons}\label{sec:baryons}

\subsection{Quantifying the bias from baryons}

The lensed CMB power spectra are sensitive to non-linearities in the CMB lensing potential. On small scales, poorly understood ``baryonic'' processes, caused by the behavior of complex visible (``baryonic'') matter (such as AGN feedback and gas cooling) can cause a non-negligible  suppression in the matter power spectrum $P_m(k)$, and therefore the lensing power spectrum $C_L^{\kappa\kappa}$; this lends uncertainties to the modeling which propagate to the lensed CMB power spectra.

As we do not have first-principles calculations of these effects, much of our current understanding comes from performing large numerical simulations, and comparing runs with and without baryonic effects included, to measure the power spectrum suppression. In this section we explore the bias induced by \emph{not} incorporating these effects into the analysis of the lensed CMB.

\begin{figure*}[t]
\includegraphics[width=\columnwidth]{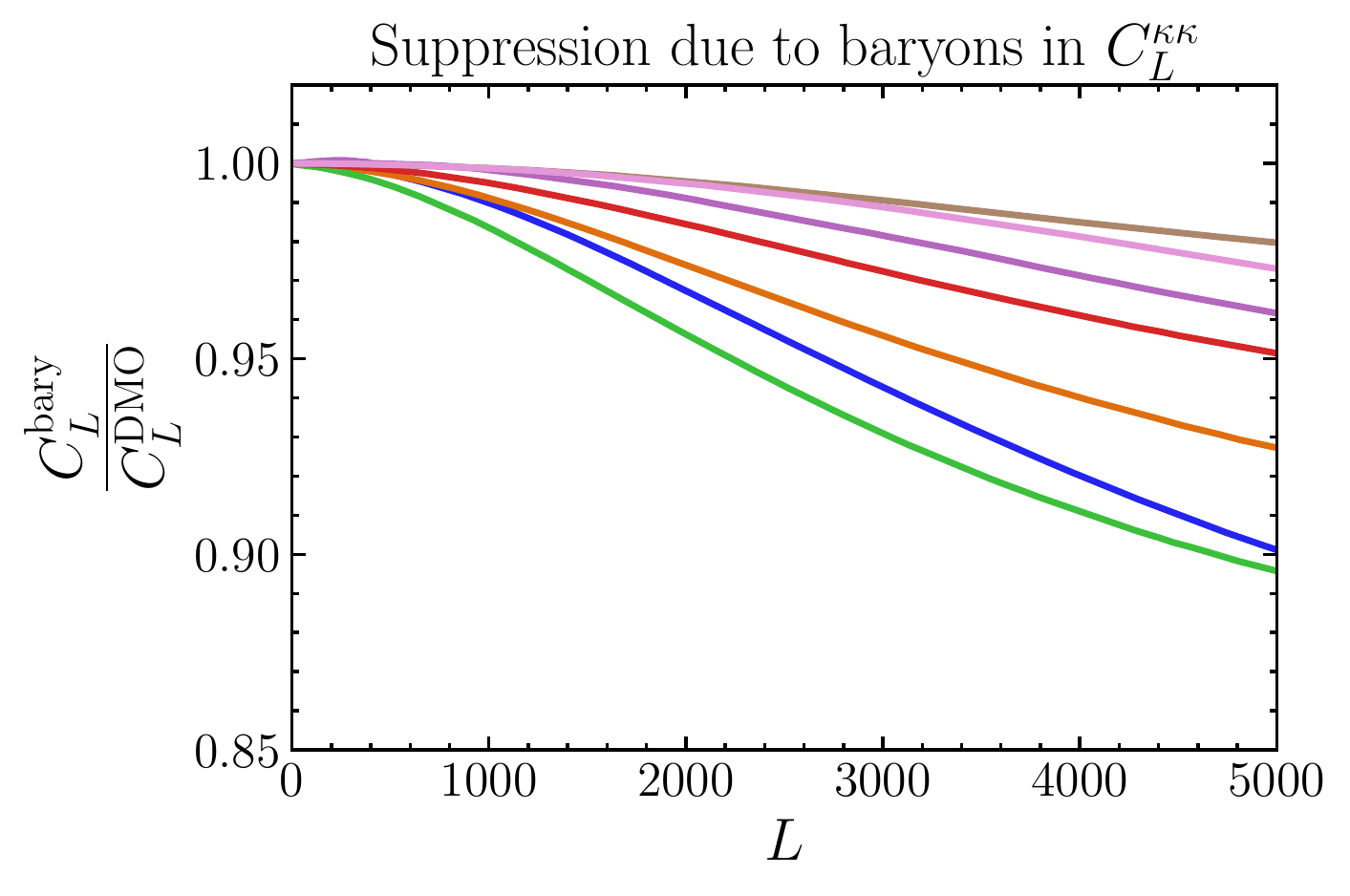}
\includegraphics[width=\columnwidth]{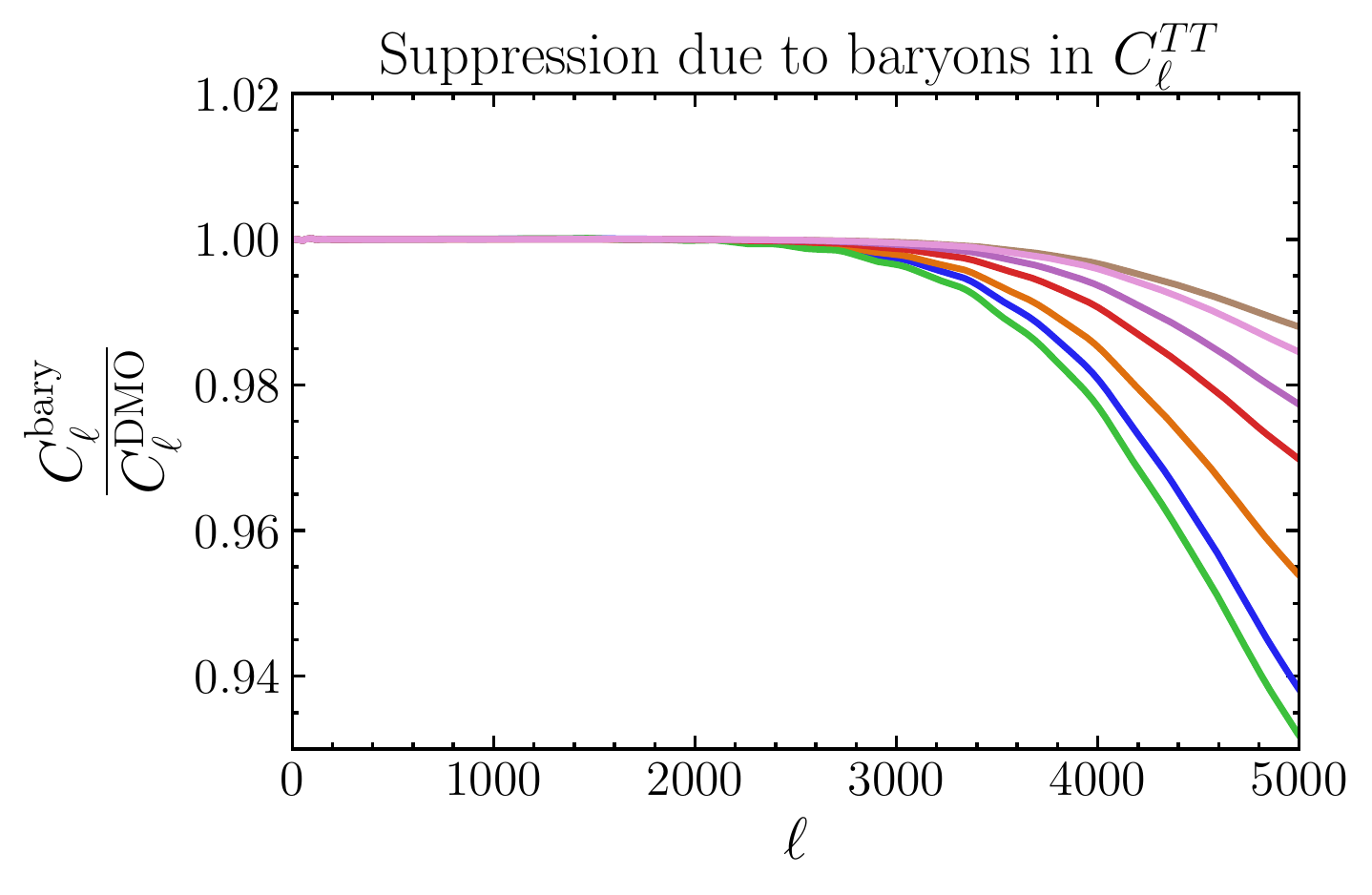}
\includegraphics[width=\columnwidth]{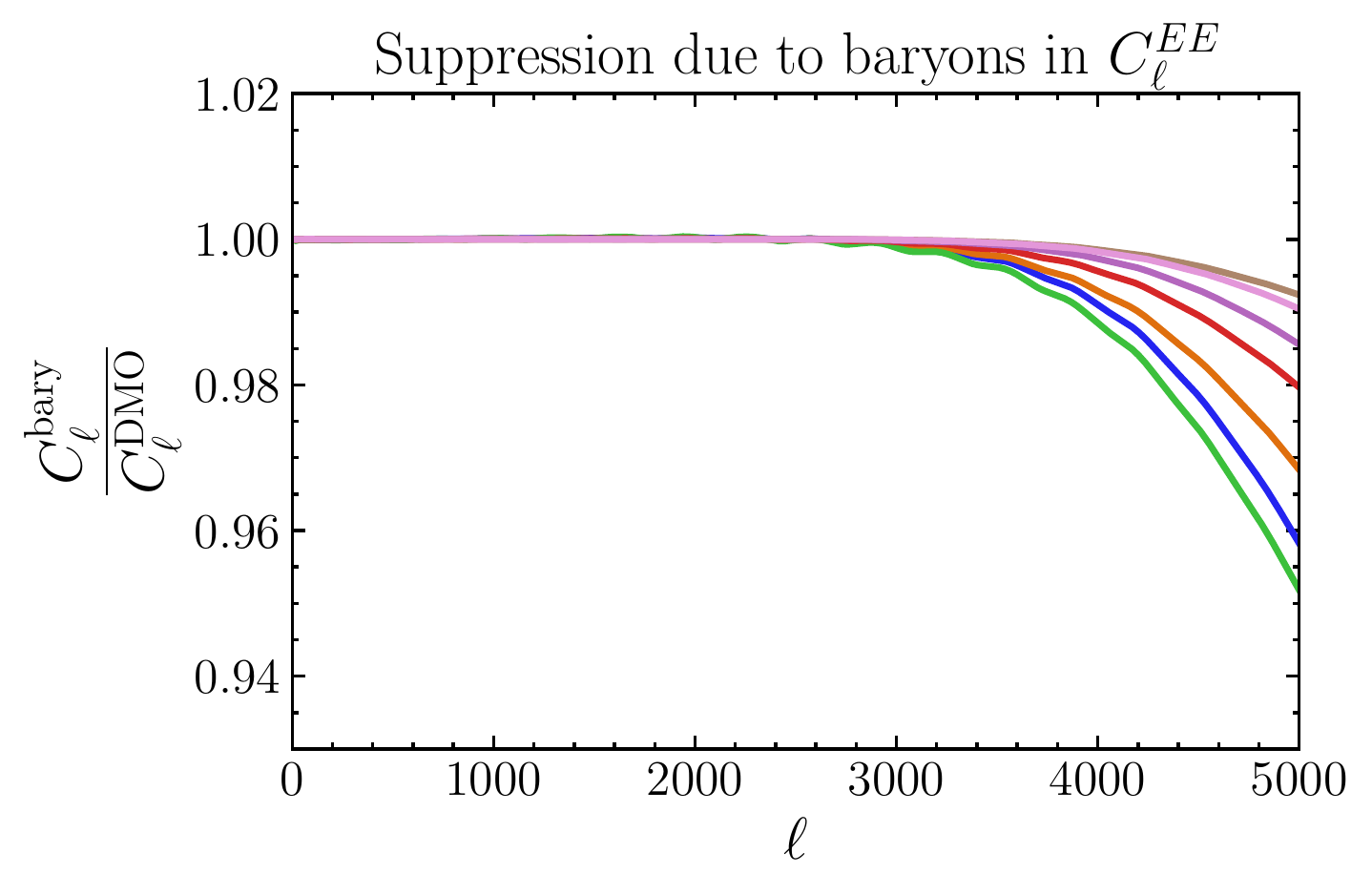}
\includegraphics[width=\columnwidth]{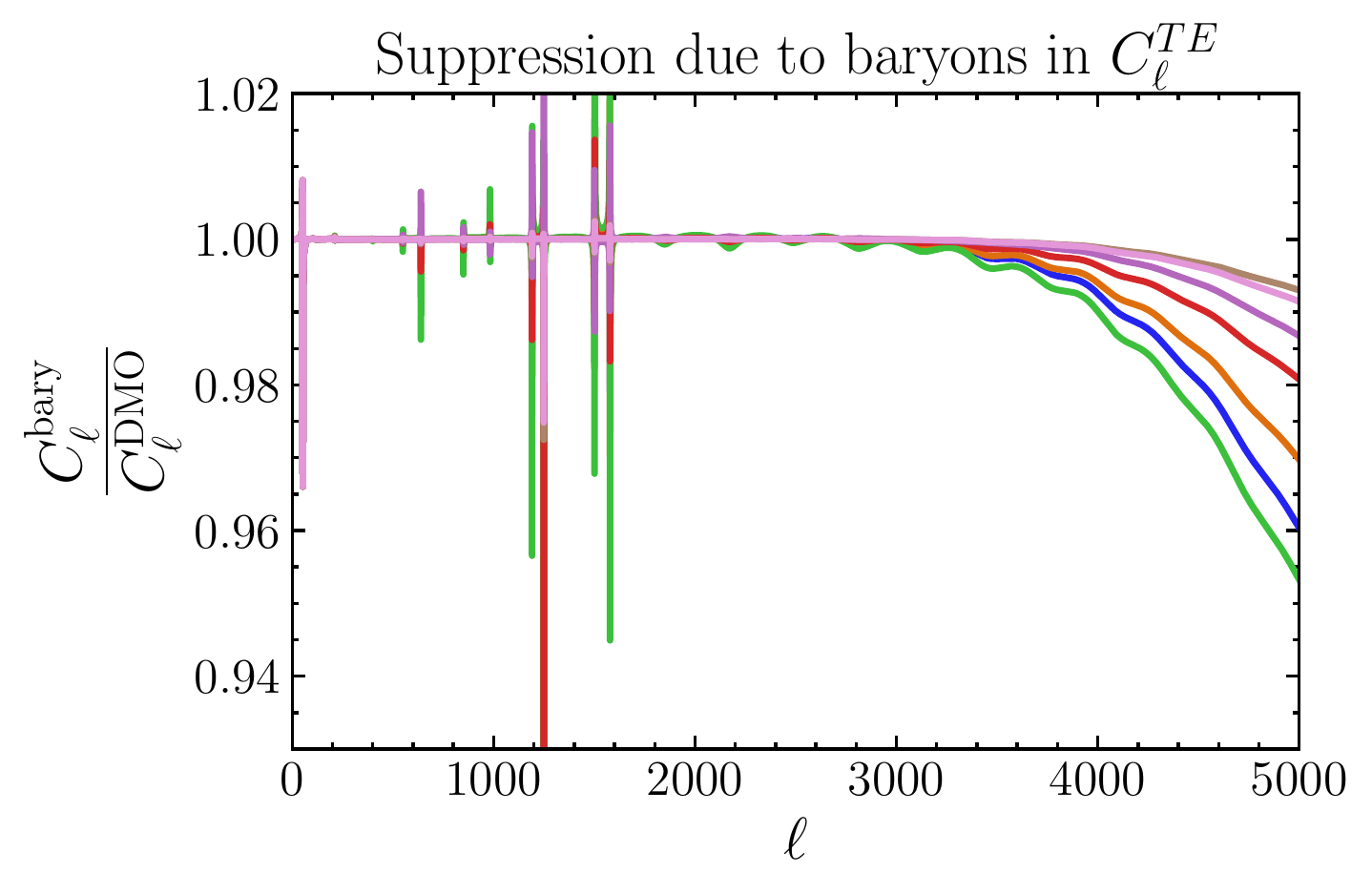}
\includegraphics[width=\columnwidth]{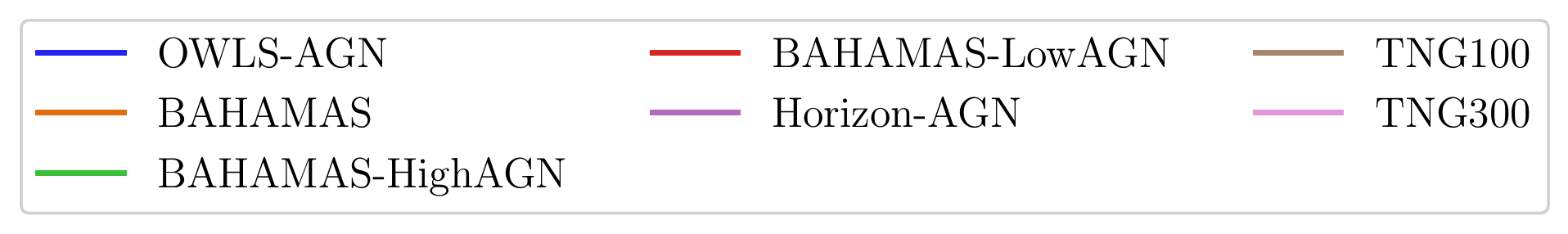}
\caption{Suppression in power due to baryonic feedback effects in various different hydrodynamical simulations. The top left panel shows the suppression in the CMB lensing power spectrum relative to a (non-linear) dark-matter-only calculation.  The other panels show the resulting impact on the lensed primary CMB power spectra ($TT$/$TE$/$EE$), as labeled.}\label{fig:baryon_suppression_allsims}
\end{figure*}

In Figure~\ref{fig:baryon_suppression_allsims}, we show the suppression in $C_L^{\kappa\kappa}$ induced by baryonic feedback effects in various hydrodynamical simulations~\cite{2020PhRvD.101f3534C,2020arXiv201106582M}.\footnote{The CMB lensing power spectrum suppression calculations are available at~\url{https://github.com/sjforeman/cmblensing_baryons}.}  The figure also shows how the baryonic suppression in the CMB lensing power spectrum propagates to suppression in the lensed primary CMB power spectra.   We show results for the OWLS-AGN simulation~\cite{vanDaalen2011,vanDaalen:2019pst}, the fiducial BAHAMAS simulation as well as its ``low-AGN'' and ``high-AGN'' variants~\cite{vanDaalen:2019pst,2017MNRAS.465.2936M,2018MNRAS.476.2999M}, the Horizon-AGN simulation~\cite{2014MNRAS.444.1453D,2016MNRAS.463.3948D,2018MNRAS.480.3962C}, and the TNG100 and TNG300 runs from the Illustris-TNG simulation suite~\cite{2018MNRAS.475..648P,2018MNRAS.475..676S,2018MNRAS.475..624N,2018MNRAS.477.1206N,2018MNRAS.480.5113M,2019ComAC...6....2N}.  All of the simulations yield qualitatively similar predictions for the power suppression due to baryonic feedback on the range of scales of interest here; however, the exact shape and amplitude of the suppression varies depending on the exact implementation (e.g., comparing the three BAHAMAS runs, one can see that the stronger the feedback prescription, the larger the predicted suppression, as gas is blown further out of halos into the intergalactic medium).  We consider all eight of these baryonic models in the following, although our tabulated numerical results (e.g.,~Table~\ref{tab:biases_SOS4}) will generally focus on the OWLS-AGN run from the OWLS simulation suite~\cite{vanDaalen2011,vanDaalen:2019pst}.

The power \textit{suppression} can be computed from each simulation by comparing the power spectrum from a ``dark-matter-only'' (DMO) run with the full baryonic physics (``AGN'') run. The measured power spectrum ratio,
\be
\hat R(k,z) \equiv \frac{\hat P^{\mathrm{AGN}}(k,z)}{\hat P^{\mathrm{DMO}}(k,z)},
\ee
can be constrained from the simulations much better than either of the power spectra $\hat P^{\mathrm{DMO},\mathrm{AGN}}(k,z)$ directly, as much of the sample variance in the measurements of the separate power spectra cancel in their ratio. We can then incorporate baryons into the non-linear matter power spectrum by modifying a theoretically calculated $P^{\mathrm{DMO}}(k,z)$ according to
\be
P^{\mathrm{bary}}(k,z) = \hat R(k,z) P^{\mathrm{DMO}}(k,z).
\ee
While $\hat R(k,z)$ has some dependence on the cosmology 
\cite{vanDaalen:2019pst} we do not consider this effect here; this is sufficient for our forecasts, particularly as derivatives with respect to power spectra computed with $\hat R(k,z)$ are never computed.

By using $P^{\mathrm {bary}}(k,z)$ in Eq.~\eqref{clkappakappa}, we compute the CMB lensing power spectrum incorporating baryonic feedback, $C_L^{\kappa\kappa}{}^{\mathrm{bary}}$. We then use the pyCAMB function \texttt{get\_lensed\_cls\_with\_spectrum(clkk)} to obtain the $C^{TT,TE,EE}_\ell{}^{\mathrm{bary}}$ from $C_L^{\kappa\kappa}{}^{\mathrm{bary}}$.\footnote{We have explicitly checked that the output of this function when using $C_L^{\kappa\kappa}{}^{\rm DMO}$ agrees with the usual CAMB output for the lensed CMB power spectra, i.e., the DM-only model here matches that in fiducial CAMB calculations.}  We then define our $\Delta C_\ell$ to be used in Eq.~\eqref{bias},
\be
\Delta C_\ell \equiv C_\ell^{\mathrm{bary}}-C_\ell^{\mathrm{DMO}},
\ee
where $C_\ell{}^{\mathrm{DMO}}$ is the fiducial $C_\ell^{TT,TE,EE}$ computed with $C_L^{\kappa\kappa}{}^{\rm DMO}$. Note that the Fisher matrix (Eq.~\eqref{fisher_matrix}) in this section is always computed by taking derivatives of $C_\ell^{\rm DMO}$, as this is the fiducial model that would be used in data analysis in this scenario (and indeed, this is the fiducial model used in standard CMB analyses to date)\footnote{Note that the default HMCode setting in CAMB, as of writing, is a DM-only model.}. 

\begin{figure}[!t]
\includegraphics[width=\columnwidth]{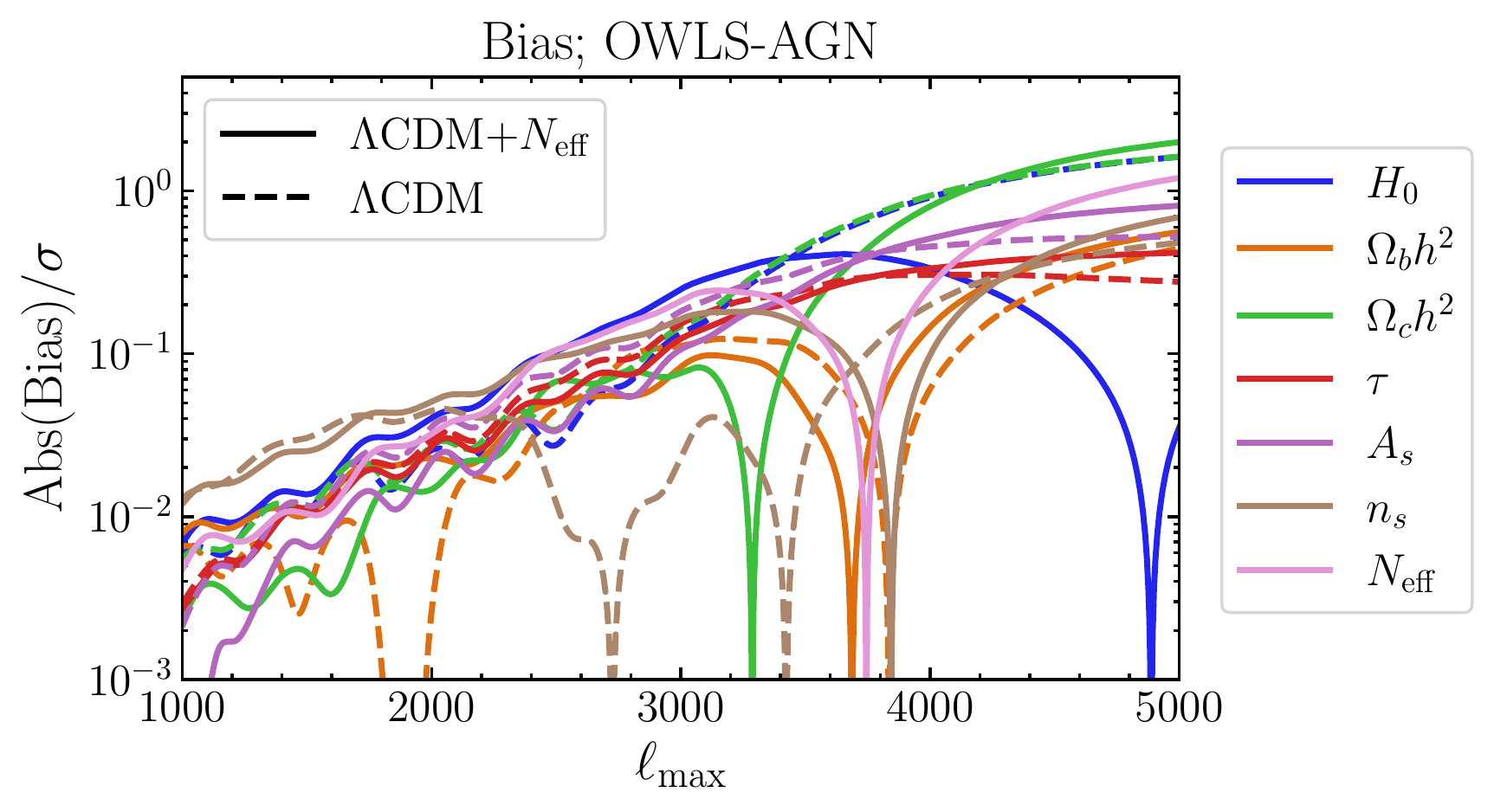}
\caption{Baryonic-feedback-induced fractional biases (in absolute value) on the inferences of parameters in the $\Lambda$CDM and $\Lambda$CDM+$\Neff$ models, as a fraction of the forecast $1\sigma$ constraints (which are shown in Figure~\ref{fig:constraints_LCDM}), plotted against the maximum multipole included in the analysis, $\ell_{\mathrm{max}}$.  Results are shown specifically for CMB-S4 here (see Table~\ref{tab:biases_SOS4} for numerical results, including for SO). The baryonic model used here is from the OWLS-AGN simulation.}
\label{fig:baryon_bias}
\end{figure}

Figure~\ref{fig:baryon_bias} presents a key result of this work.  In this plot, we show the fractional biases (in units of $\sigma$) on the parameters in the $\Lambda$CDM and $\Lambda$CDM+$N_{\rm eff}$ models induced by neglecting the baryonic suppression in $C_L^{\kappa\kappa}$, using the method described above.  We use the OWLS-AGN baryonic model here, and show results for the CMB-S4 post-ILC noise power spectra presented in Figure~\ref{fig:lensed_unlensed_power}.  The biases are presented as a function of $\ell_{\rm max}$, the maximum multipole considered in the analysis of the $TT$, $TE$, and $EE$ power spectra ($\ell_{\rm max}$ is taken to be the same for all three spectra here).  At high $\ell_{\mathrm{max}}$ the biases become increasingly significant, with some parameter biases exceeding their forecast statistical error bar.   For example, with $\ell_{\rm max} = 5000$, the bias on $H_0$ reaches 1.6$\sigma$ in $\Lambda$CDM, while the bias on $\Neff$ reaches 1.2$\sigma$ in $\Lambda$CDM+$N_{\rm eff}$.  The exact numerical values are given in Table~\ref{tab:biases_SOS4} for $\ell_{\rm max} = 5000$, along with analogous results computed for SO.  These results clearly illustrate that these biases are potentially significant for upcoming high-precision, high-resolution CMB experiments.

\subsection{Strategies to mitigate the baryonic biases}
\subsubsection{Removing small-scale $TT$ information}

If unaccounted for, baryonic feedback effects will bias cosmological parameter inference from the primary CMB; these systematic effects will require mitigation.  A simple approach is to impose a lower $\ell_{\mathrm{max}}$ cut than 5000 on the data used for the analysis; it is clear from Figure~\ref{fig:constraints_LCDM} that there is not much constraining power at $\ell\greaterthanapprox3000$, while in Figure \ref{fig:baryon_bias} we see that the biases increase significantly on these scales. Indeed, if one removes \textit{only} the $TT$ information at $\ell > 3000$, while keeping the $TE$ and $EE$ spectra in the analysis, the bias is significantly reduced, as we now show.

\begin{figure}[!t]
\includegraphics[width=\columnwidth]{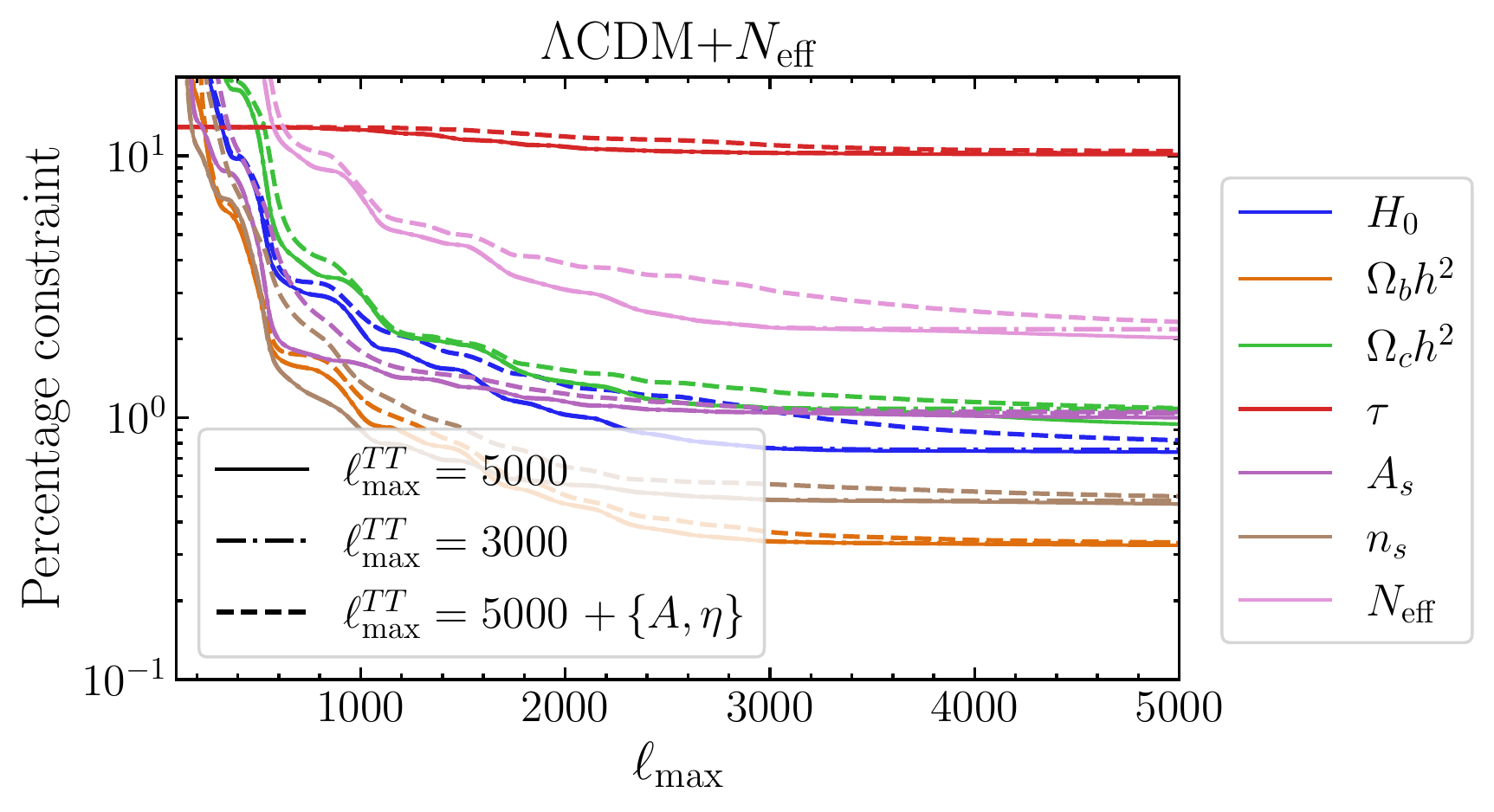}
\caption{The impact of the various mitigation methods --- a strict $\ell_{\mathrm{max}}=3000$ cut on $TT$ and marginalization over baryonic feedback parameters $(A,\eta)$ --- on the constraints in the $\Lambda$CDM$+\Neff$ model. The errors are only marginally increased in both cases, with the biggest increase seen in $\Omega_ch^2$; similar conclusions hold for the $\Lambda$CDM model analysis.  Note that the horizontal axis here has the same meaning as in Figure~\ref{fig:baryon_bias}, but that $\ell_{\rm max}^{TT}$ is not increased above 3000 in the case shown in the dash-dotted curves.}\label{fig:ttmax3000_constraints}
\end{figure}

\begin{figure*}[!t]
\includegraphics[width=0.32\textwidth]{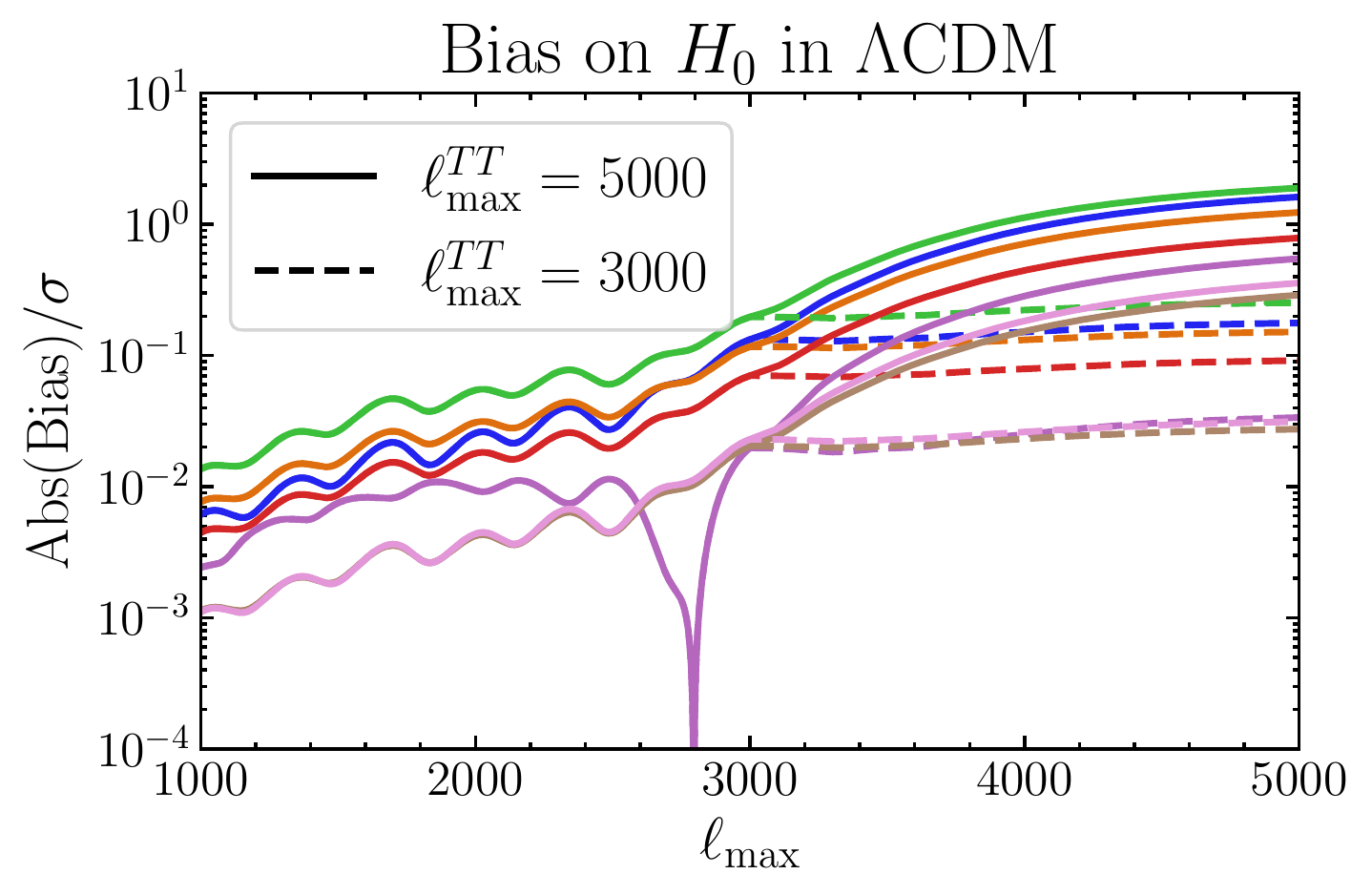}
\includegraphics[width=0.32\textwidth]{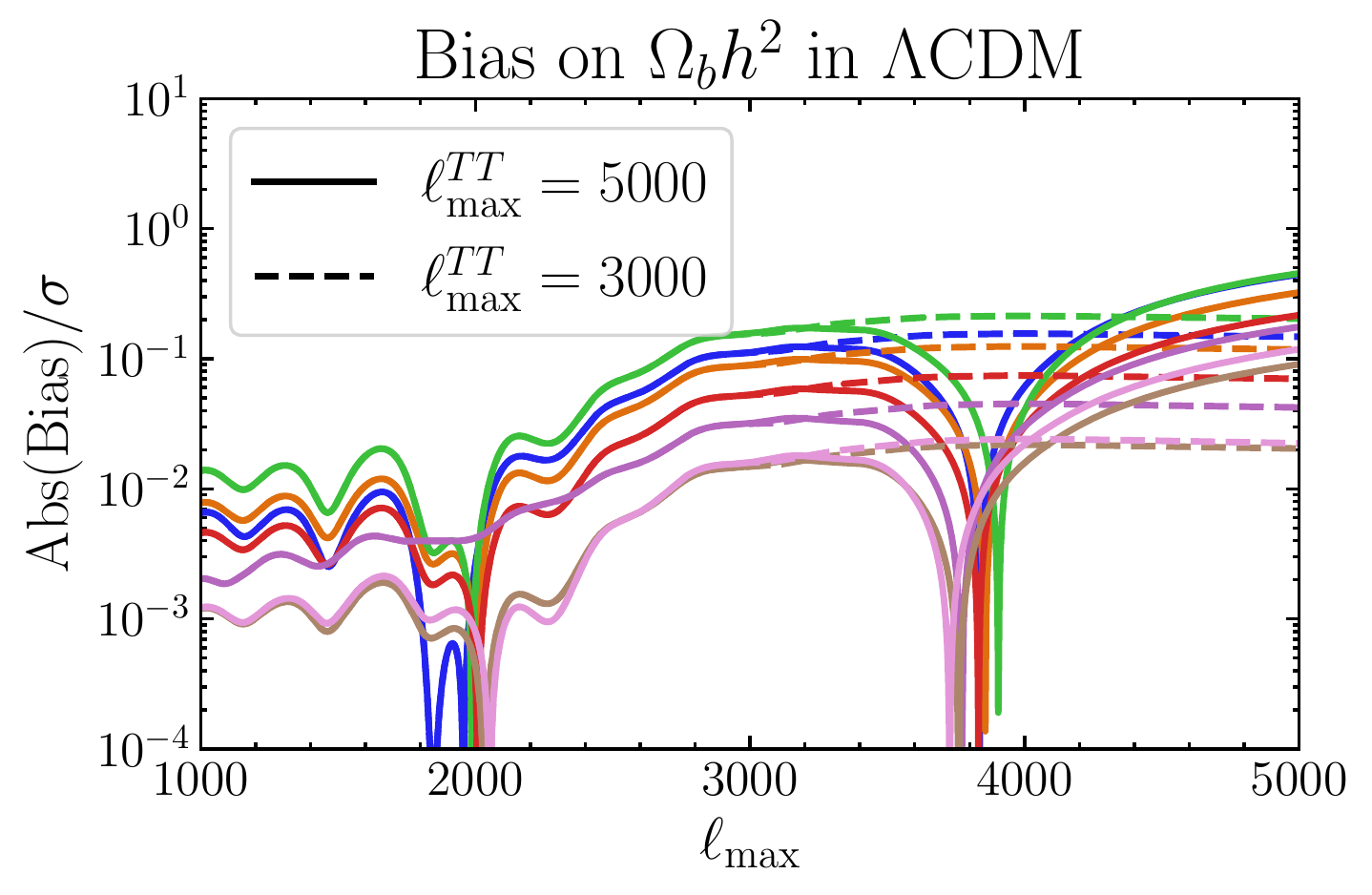}
\includegraphics[width=0.32\textwidth]{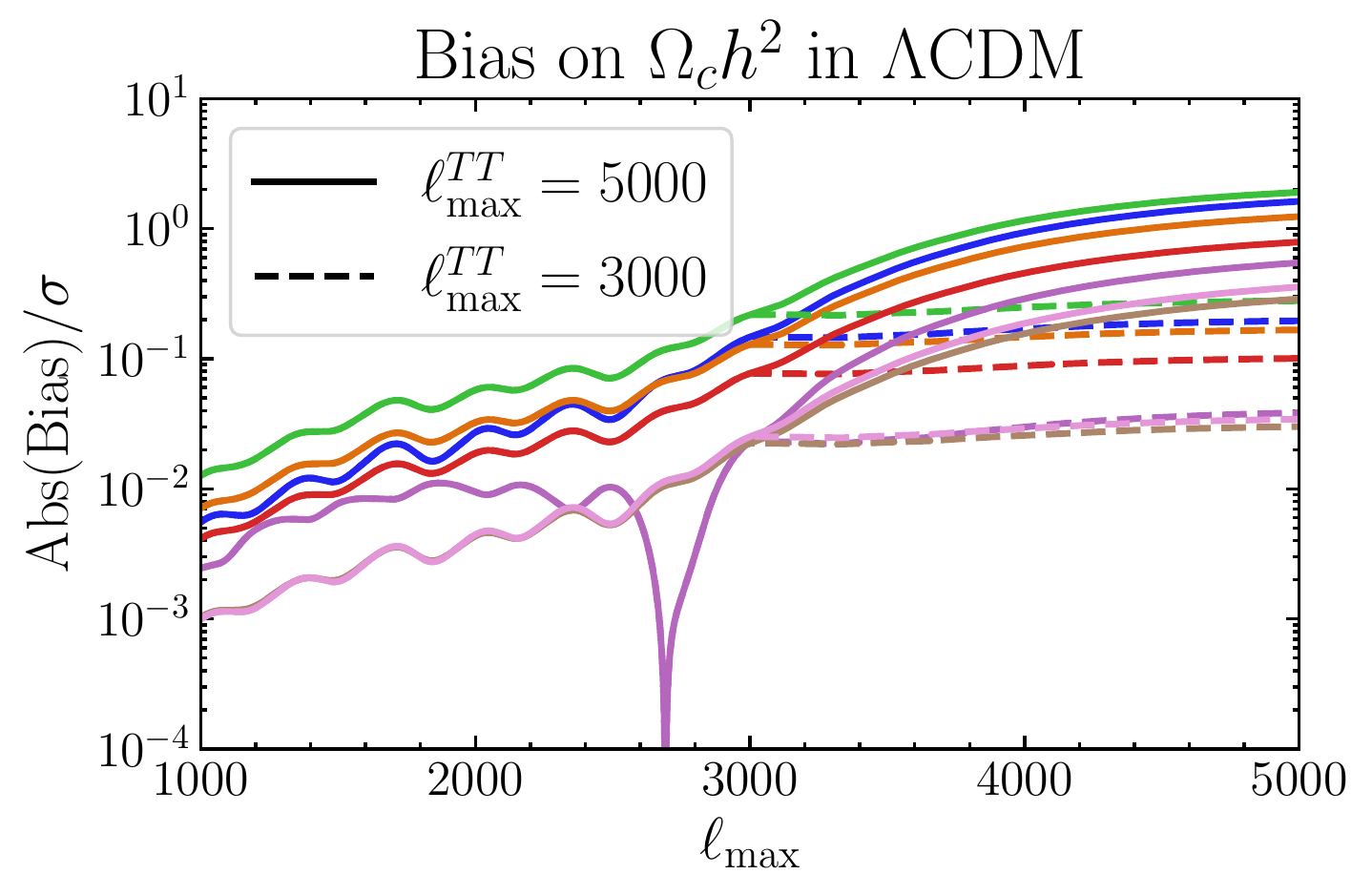}
\includegraphics[width=0.32\textwidth]{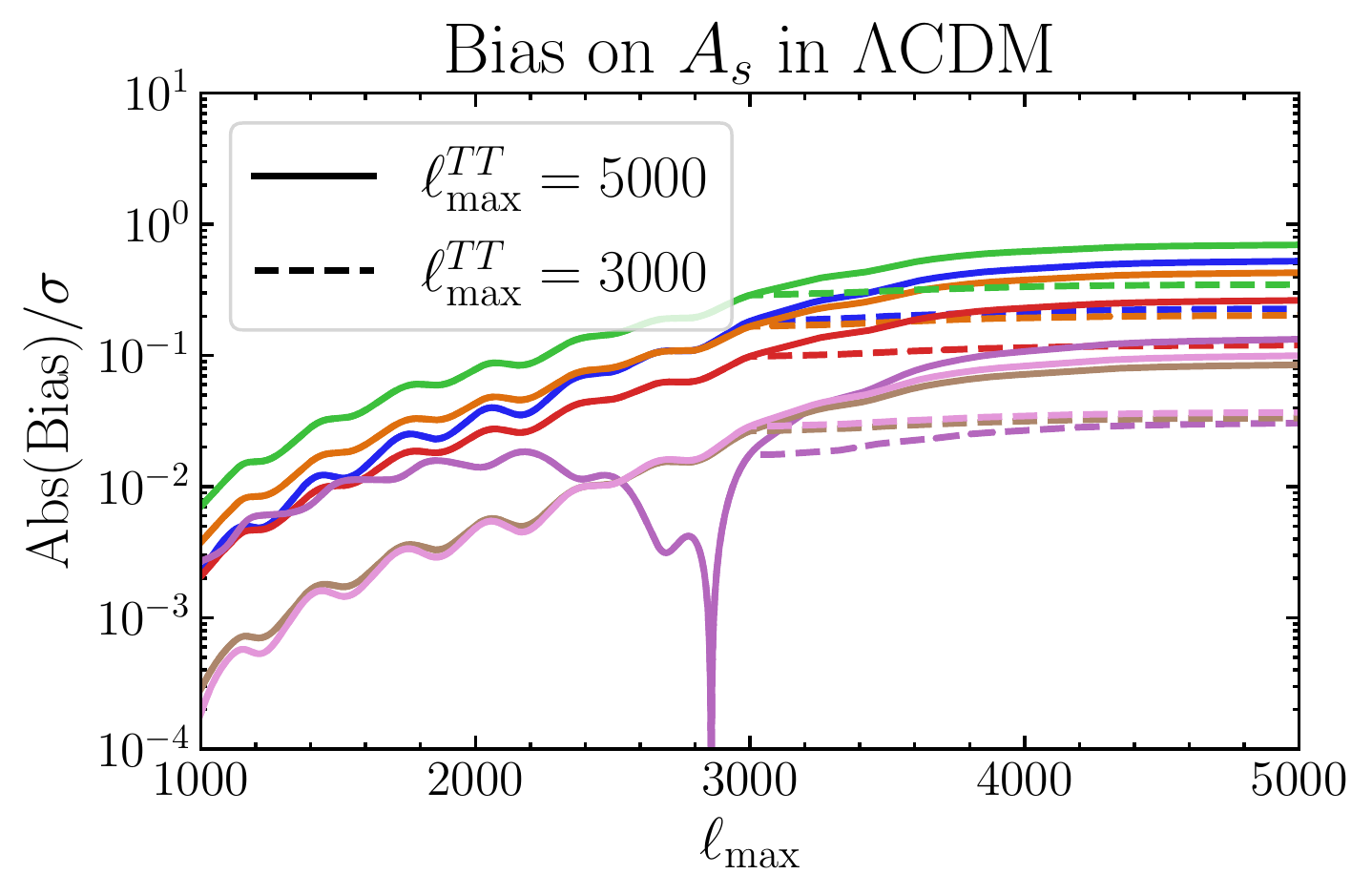}
\includegraphics[width=0.32\textwidth]{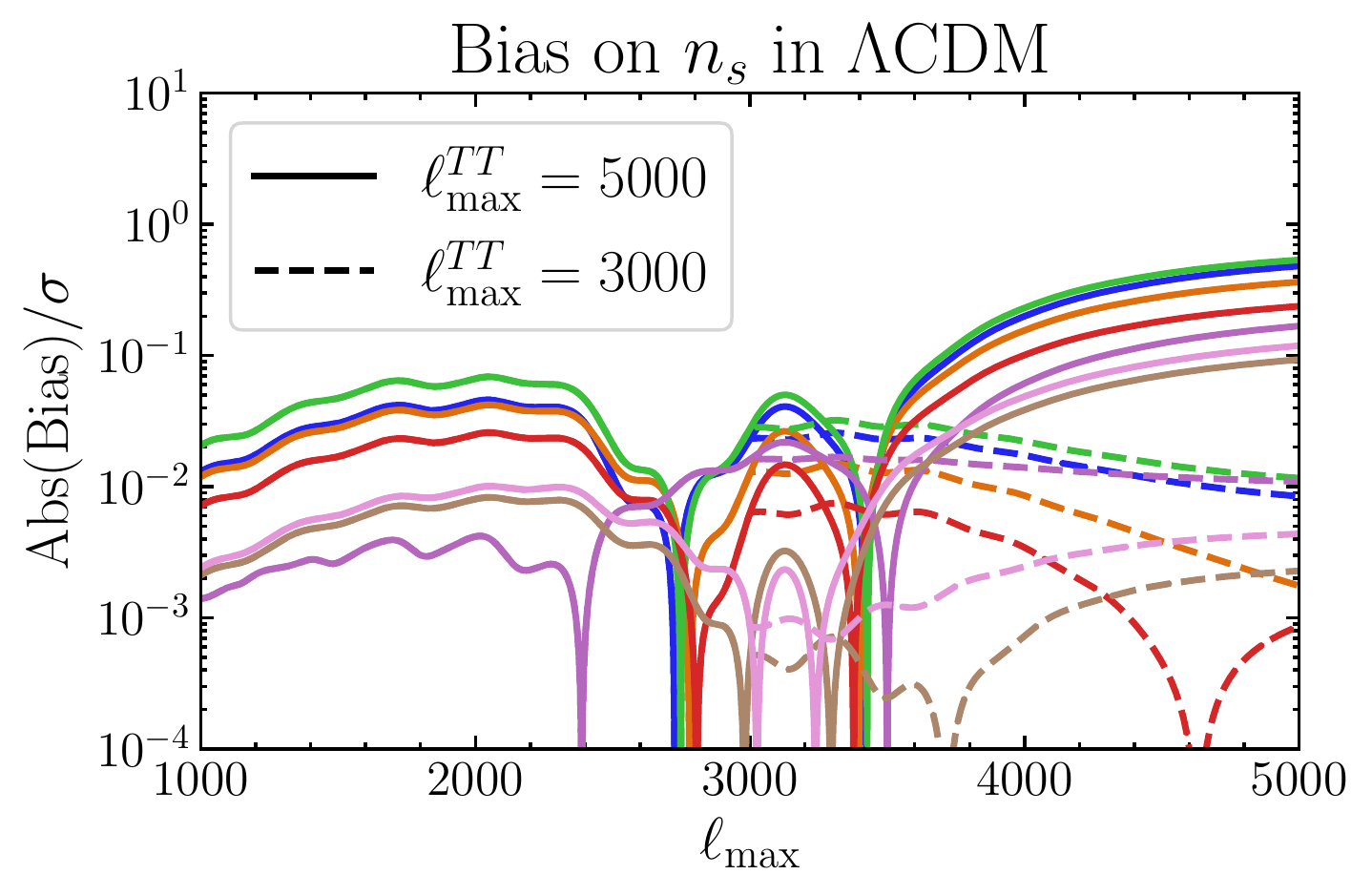}
\includegraphics[width=0.32\textwidth]{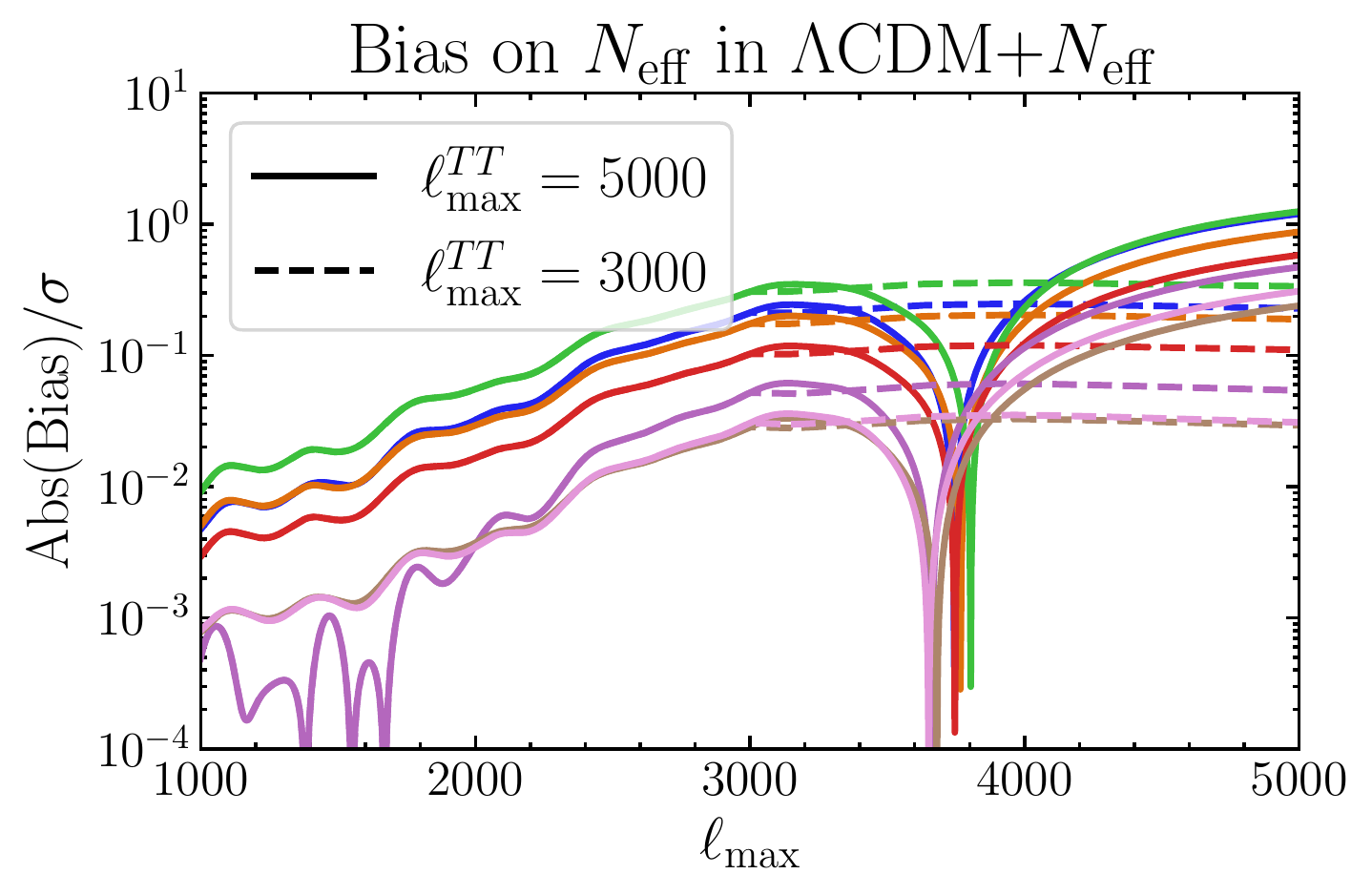}
\includegraphics[width=\columnwidth]{Images/legend_baryons.pdf}
\caption{The $\ell_{\mathrm{max}}$ dependence of the biases for each parameter in the $\Lambda$CDM model, and also for $\Neff$ in the $\Lambda$CDM$+\Neff$ model, for the different simulations shown in  Figure \ref{fig:baryon_suppression_allsims}.  The solid curves show the unmitigated biases (as in Figure~\ref{fig:baryon_bias} for OWLS-AGN), while the dashed curves show the results after mitigating the biases by imposing an $\ell_{\mathrm{max}}^{TT}=3000$ cut. All results here are computed for CMB-S4.}\label{fig:lmaxcut_allsims}
\end{figure*}

The forecast constraints with $TT$ data discarded at $\ell>3000$ are shown in Figure~\ref{fig:ttmax3000_constraints} in the dash-dotted curves for the $\Lambda$CDM$+\Neff$ model (the results for the $\Lambda$CDM model are similar).  For comparison, the previous case where $TT$ data are included up to $\ell=5000$ is shown in the solid curves.  While it is clear that the marginalized parameter error bars increase somewhat, the overall penalty is generally mild ($\lesssim 5$\%, except for $\Omega_c h^2$, which is impacted somewhat more than this).  Numerical results for this increase in error bars are collected in Appendix~\ref{app:biases} in Tables~\ref{tab:LCDM_forecast_full} ($\Lambda$CDM) and~\ref{tab:LCDMNeff_forecast_full} ($\Lambda$CDM+$\Neff$).

The parameter biases in this case (when no $TT$ information is considered above $\ell=3000$) are shown in Figure~\ref{fig:lmaxcut_allsims} for all eight baryonic physics models. It is clear from the flattening of the bias curves at $\ell\gtrsim3000$, above which $TT$ data are excluded, that most of the bias is incurred from $C_\ell^{TT}$ in this small-scale regime.  Thus, the baryonic biases can be controlled with a strict $\ell_{\mathrm{max}}$ cut on $TT$.  Numerical results for this approach, analogous to those in Table~\ref{tab:biases_SOS4}, are collected in Table~\ref{tab:biases_SOS4_lmax3000} (for the OWLS-AGN simulation) in Appendix~\ref{app:biases}. 

There are two primary reasons why the baryonic biases are dominated by the $TT$ power spectrum: (1) SO and CMB-S4 will measure more signal-dominated modes in temperature than in polarization (see the noise curves in Figure~\ref{fig:lensed_unlensed_power}); (2) at a given multipole in the damping tail, the fractional contribution of lensing to the total power is larger in $TT$ than in $TE$ or $EE$, due to the larger gradient in the unlensed temperature field.  Thus, since the high-$\ell$ $TT$ data are more sensitive to lensing (at fixed $\ell$), and a greater number of such modes are measured in temperature than in polarization, the baryonic feedback biases that enter via CMB lensing will be dominated by their effects on $TT$.  Explicitly discarding the $TT$ data on small scales is thus a simple and relatively robust approach to mitigate these biases.

In fact, this approach is consistent with the methodology often used in forecasts to account for the presence of extragalactic foregrounds at $\ell>3000$ in $TT$, i.e., the data in this region are frequently assumed to be unusable for primary CMB science.  This approach was used for the SO forecasting analysis, which set $\ell_{\rm max}^{TT} = 3000$, $\ell_{\rm max}^{TE} = 5000$, and $\ell_{\rm max}^{EE} = 5000$ (see Sec.~4 of Ref.~\cite{Ade:2018sbj}).      However, for the CMB-S4 forecasting analysis, it was assumed that $\ell_{\rm max} = 5000$ for all of the spectra, including $TT$ (see Sec.~A.2.4 of Ref.~\cite{CMBS4DSR}).  In addition, the Atacama Cosmology Telescope (ACT) DR4 CMB likelihood considers $TT$/$TE$/$EE$ data to $\ell_{\rm max} = 4325$ for all of the spectra (see Table 18 of Ref.~\cite{Choi:2020ccd}).  Our results provide motivation to explicitly discard the $TT$ data at $\ell>3000$ when performing primary CMB data analysis.

\subsubsection{Marginalizing over a model for baryons}
\label{sec:baryonmarg}

\begin{figure*}[!t]
\includegraphics[width=0.32\textwidth]{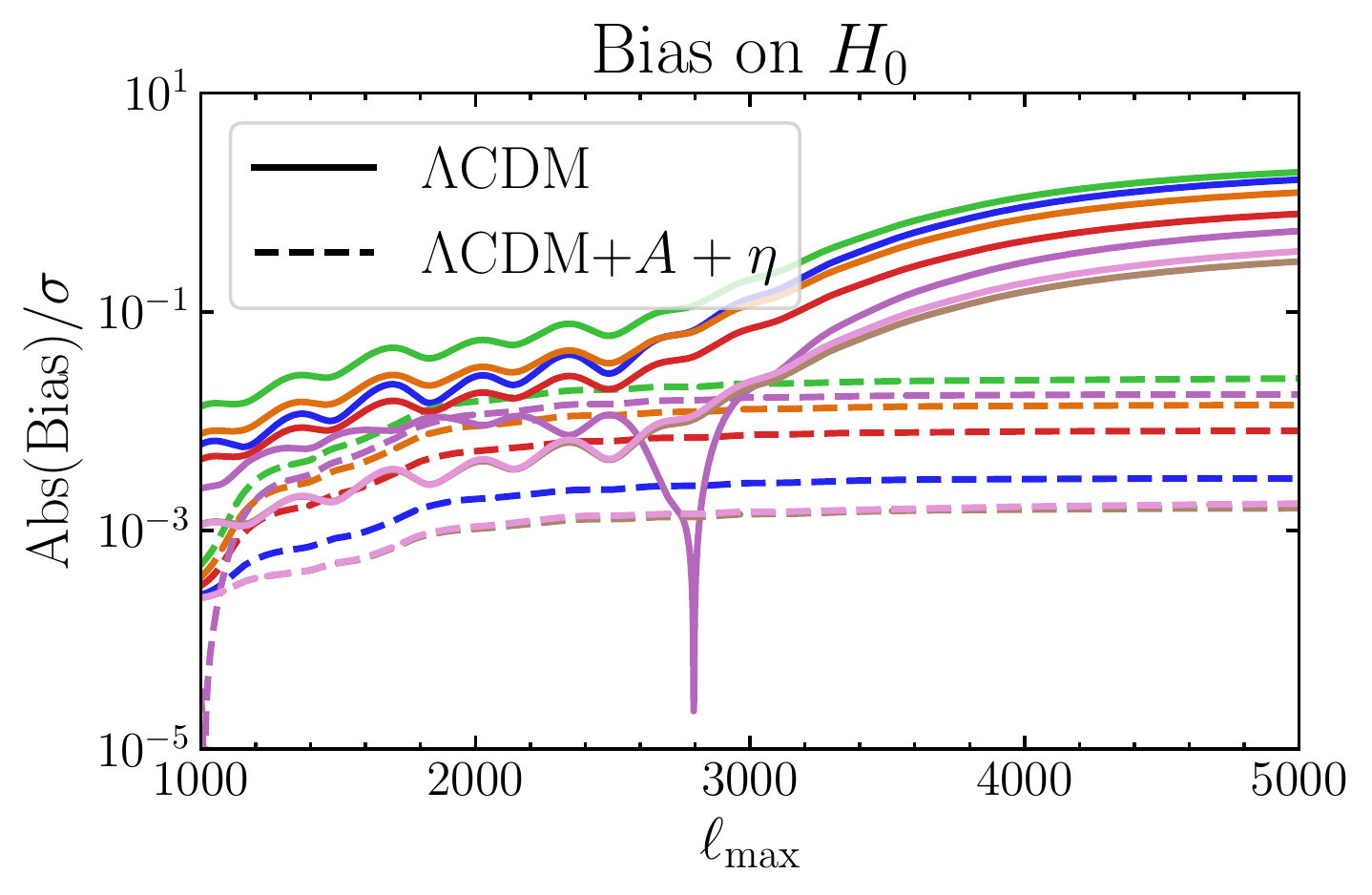}
\includegraphics[width=0.32\textwidth]{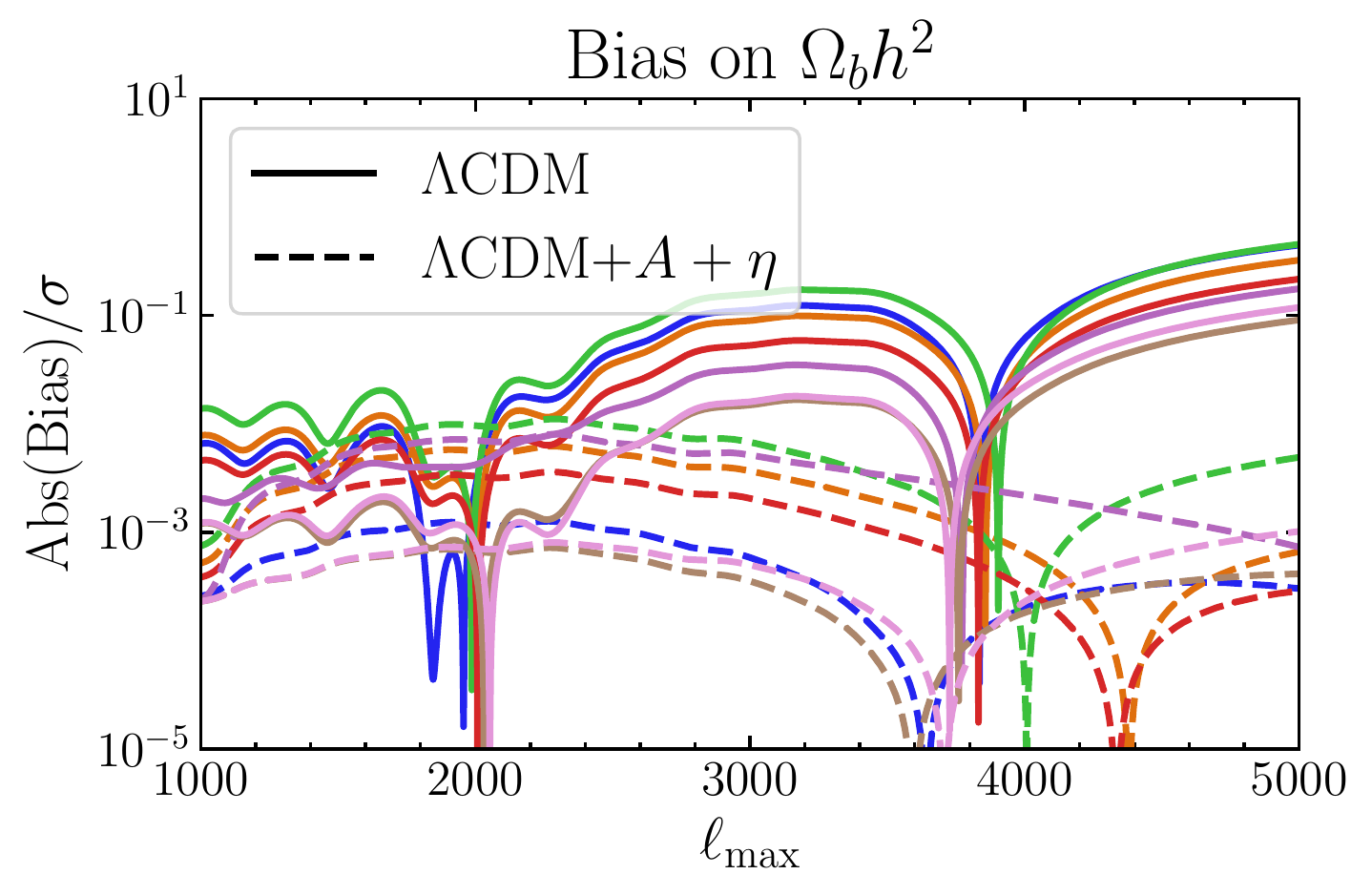}
\includegraphics[width=0.32\textwidth]{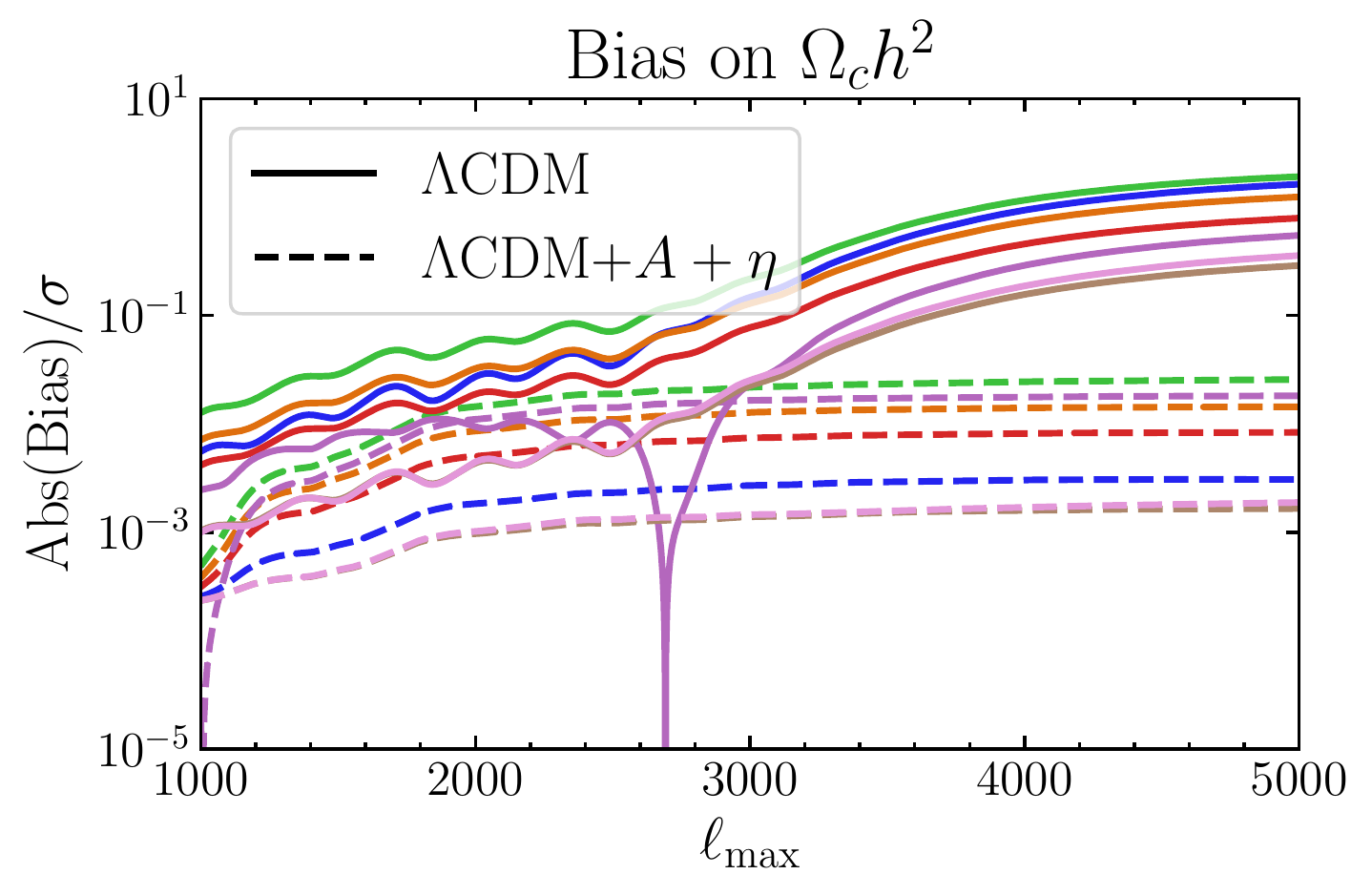}
\includegraphics[width=0.32\textwidth]{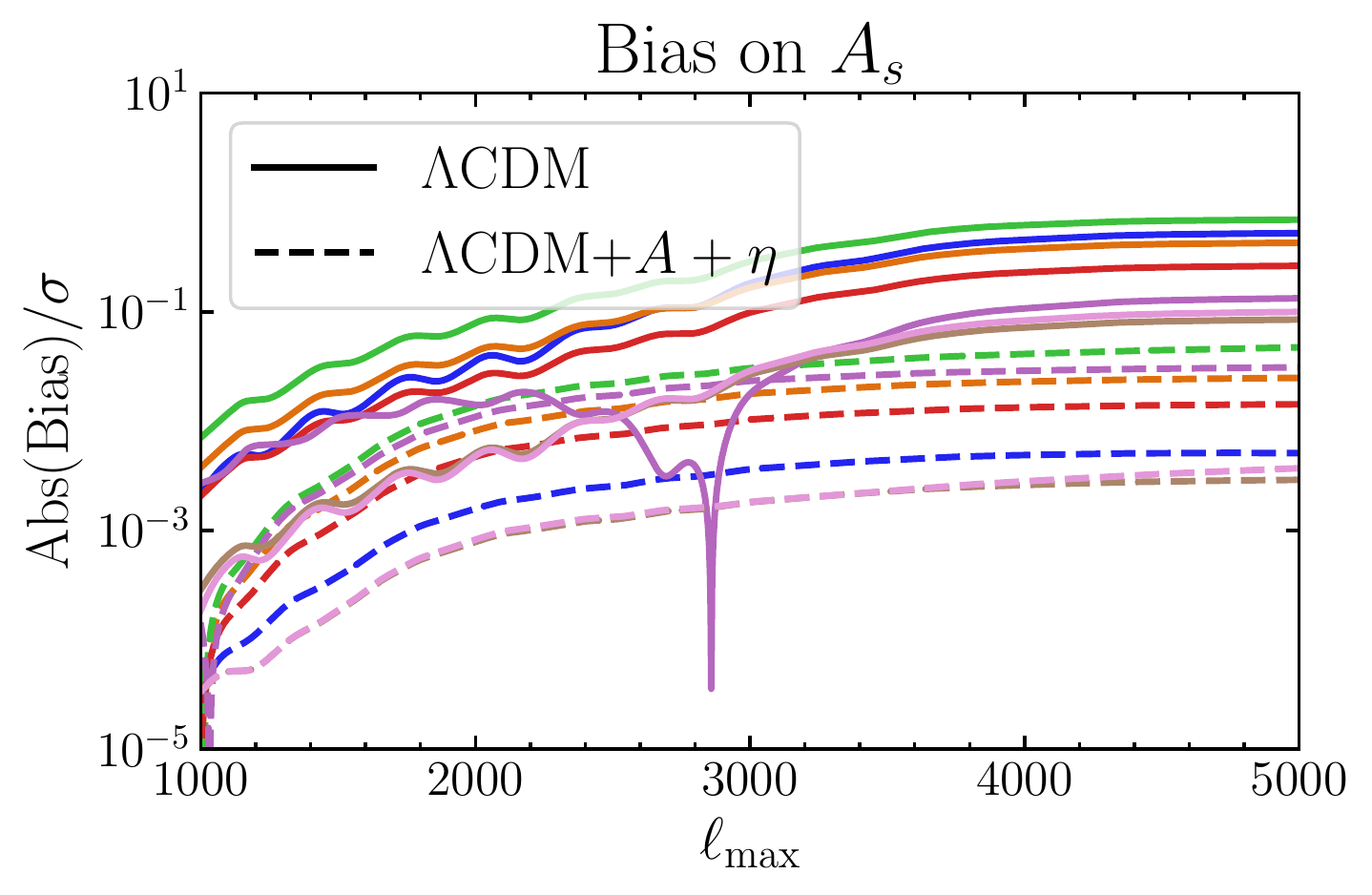}
\includegraphics[width=0.32\textwidth]{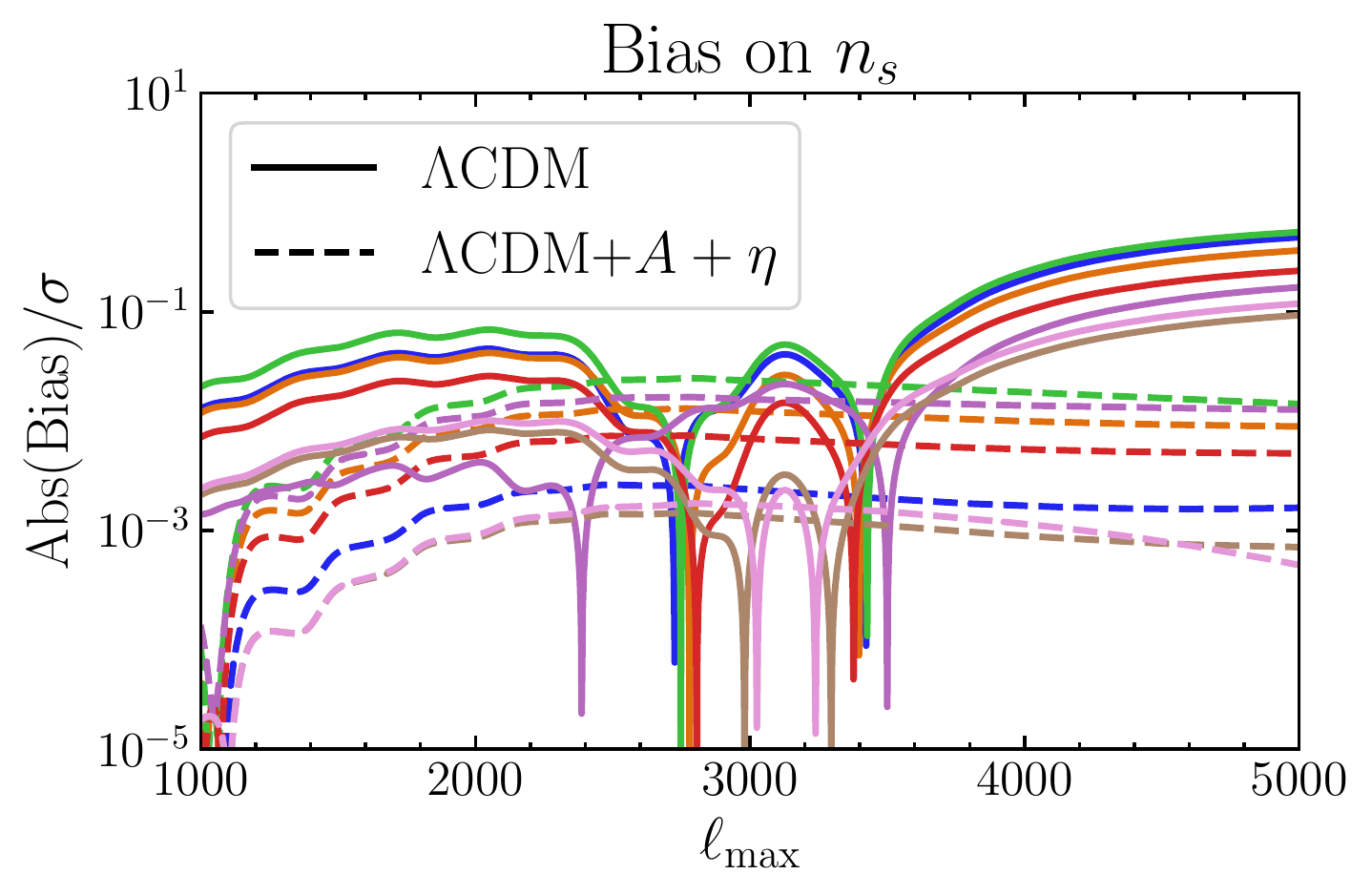}
\includegraphics[width=0.32\textwidth]{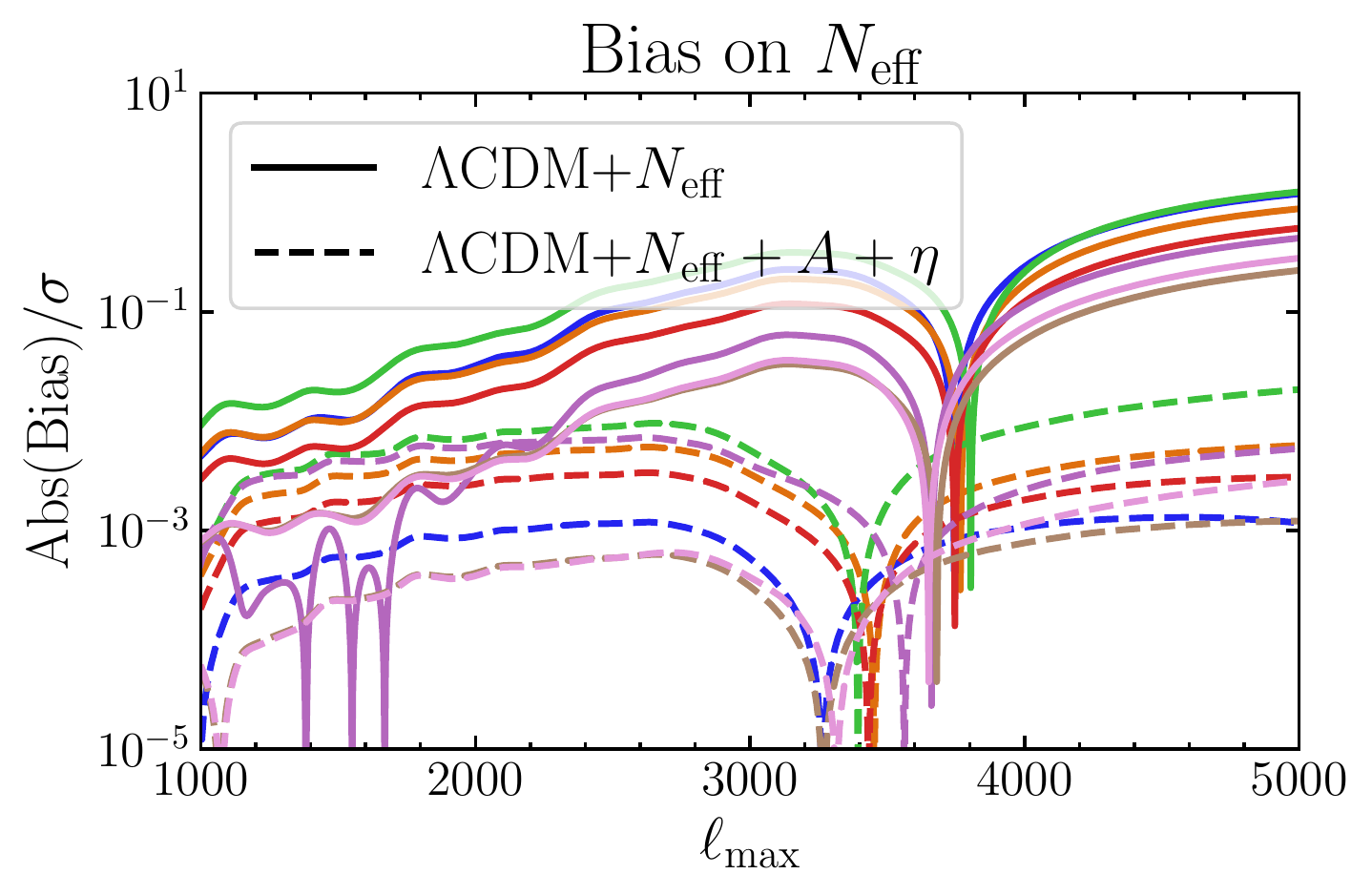}
\includegraphics[width=\columnwidth]{Images/legend_baryons.pdf}
\caption{The $\ell_{\mathrm{max}}$ dependence of the biases for each parameter in the $\Lambda$CDM model, computed for the different hydrodynamical simulations shown in  Figure \ref{fig:baryon_suppression_allsims}.  The dashed curves show the unmitigated biases (as in Figure~\ref{fig:baryon_bias} for OWLS-AGN), while the solid curves show the results after mitigating the biases by marginalizing over baryonic feedback parameters $(A,\eta)$. We also include (bottom right) the biases on $\Neff$ in the $\Lambda$CDM$+\Neff$ model.  All results here are computed for CMB-S4.}\label{fig:marginalisation_allsims}
\end{figure*}

An alternative way to avoid biases from mismodeling baryons is to incorporate them into the model for $C_L^{\kappa\kappa}$. While we do not know the true theory describing baryonic effects in our universe, certain (semi-)analytic models have been shown to accurately capture these effects on the matter power spectrum.  We focus specifically on the approach of Ref.~\cite{2015MNRAS.454.1958M}, in which a two-parameter model is introduced to incorporate the effects of baryons.  In particular, the non-linear matter power spectrum is modelled with a modified halo model. While we will not go into the details of the halo model here (see, e.g., \cite{Cooray:2002dia} for a comprehensive review), we note that \cite{2015MNRAS.454.1958M} extends the halo model by introducing two parameters $A$ and $\eta$, which can be varied to allow for different prescriptions of baryonic effects. $A$ modifies the halo concentration $c(M,z)$ in a parametric way:
\be
c(M,z)=A\frac{1+z_f}{1+z} \,,
\ee
where $z_f(M)$ is the formation redshift of halos of mass $M$ and redshift $z$. The parameter $\eta$ modifies the Fourier-transformed halo density profile according to 
\be
u(k,M,z)\rightarrow u(\nu^\eta k,M,z)\label{ukmz}
\ee
where $\nu=\frac{\delta_c}{\sigma(M)}$ with $\delta_c$ the critical density required for spherical collapse and $\sigma(M)$ the variance in the initial density fluctuation field when smoothed with a top-hat filter with the size of the virial radius of the halo.  $\eta$ is parametrized as
\be
\eta=\eta_0-0.3\sigma_8(z)
\ee
where $\sigma_8(z)$ is the linear-theory rms amplitude of density fluctuations over a sphere with radius $8 \, h \, \mathrm{Mpc}^{-1}$ at redshift $z$; $\eta_0$ is the  parameter that is modified to incorporate baryons. The fiducial values used for $\eta_0$ and $A$ are $\eta_0=0.603$ and $A=3.13$.

The model described above is implemented in the HMCode module within CAMB; thus, we can include the two extra parameters, $A$ and $\eta_0$, in our Fisher matrix and marginalize over them in order to mitigate the impact of baryons in a parametric manner.  Such an approach has been shown to be effective in removing the bias induced by baryons on the neutrino mass inference from the CMB lensing power spectrum, for a range of different baryon models~\cite{2020arXiv201106582M}.

Figure~\ref{fig:ttmax3000_constraints} (dashed curves) shows the effect of this marginalization on the $\Lambda$CDM+$\Neff$ parameter constraints for CMB-S4.  In contrast to the approach of discarding high-$\ell$ $TT$ data, this method increases the error bars somewhat more noticeably.  However, the penalties are still relatively mild, generally $\lesssim 10$\%.  Numerical results illustrating the increase in error bars are collected in Appendix~\ref{app:biases} in Tables~\ref{tab:LCDM_forecast_full} ($\Lambda$CDM) and~\ref{tab:LCDMNeff_forecast_full} ($\Lambda$CDM+$\Neff$).  We note that these penalties could be decreased by performing a joint analysis of the primary CMB power spectra with $C_L^{\kappa\kappa}$ inferred from the CMB four-point function; the latter observable would constrain $(A,\eta)$, thus yielding a smaller penalty when marginalizing over these parameters in the analysis.  However, a careful treatment of the joint covariance~\cite{GreenDelensing,Peloton2017} would be required, which we defer to a dedicated analysis of this method.

In Figure~\ref{fig:marginalisation_allsims}, we show the effect of this marginalization on the baryonic-feedback biases for CMB-S4 for the $\Lambda$CDM parameters and also for $\Neff$ for the full set of baryonic models; the biases generally decrease by factors of $>100$, illustrating that the marginalization is extremely effective. Numerical results for this approach (for OWLS-AGN), analogous to those in Table~\ref{tab:biases_SOS4}, are given in Table~\ref{tab:biases_SOS4_aeta} in Appendix~\ref{app:biases}.  

Overall, we conclude that this method is very promising, although one may worry that if there is a significant mismatch between the assumed parameterization and the actual baryonic effects in our universe, its effectiveness could be curtailed (note that Figure~\ref{fig:marginalisation_allsims} provides evidence against this concern).  A range of hydrodynamical simulations should be used to ensure its robustness in upcoming high-precision CMB analyses.

\subsubsection{Delensing}
\label{sec:delensing}

Delensing is the process of ``undoing'' the effects of gravitational lensing on the CMB. It requires knowledge of the actual CMB lensing potential on our sky, as derived from a reconstructed lensing map --- or proxies for it, such as the measured CIB field or galaxy surveys (e.g.,~\cite{2012JCAP...06..014S,Sherwin:2015baa,Yu:2018tem}).  In the ideal case, the ``delensed'' CMB maps will recover the \textit{unlensed} CMB temperature and polarization fields.  This procedure is very important for enabling tight constraints on the tensor-to-scalar ratio in upcoming $B$-mode surveys~\cite{CMBS4DSR}.  For $T$ and $E$-mode maps, delensing can provide slightly improved constraints on some of the parameters \cite{GreenDelensing}, but we note here that it is to be expected that it will provide \textit{unbiased} constraints on the parameters as well, as sensitivity to low-$z$ baryonic feedback effects will be reduced (or, ideally, removed).

Note that as we want to delens the high-$\ell$ power spectra here, a good proxy of the high-$L$ CMB lensing potential will be required. In particular, Eq.~\eqref{eq.ClTTlensed_highell} indicates that to delens the CMB spectra up to $\ell \approx 5000$, we will need to have knowledge of $C_L^{\phi\phi}$ also out to $L \approx 5000$.  This is a much smaller-scale regime than has been focused on in most previous delensing work, e.g., for primordial $B$-mode delensing.  The reconstructed CMB lensing potential maps from SO or CMB-S4 are unlikely to have high fidelity at $L \approx 5000$ (e.g., see Figure~6 of Ref.~\cite{Ade:2018sbj}), although improved small-scale estimators could help to some extent~\cite{2019PhRvD.100b3547H}.  Fortunately, external delensing tracers could be reasonably effective in this domain.  In particular, dense galaxy samples from LSST and other photometric surveys may present a feasible option --- see, e.g., Appendix~B of Ref.~\cite{Yu:2018tem}.  However, a dedicated study would be needed to forecast the effectiveness of small-scale delensing.

One advantage of the delensing approach is that (at least in an ideal case), a model of the baryon-affected non-linear lensing power would no longer be needed for the interpretation; if one delensed the maps perfectly, then clearly there would be no need for a model of the nonlinear lensing power to interpret the primary CMB data at all.  In a more pessimistic case in which the delensing efficiency is much less than 100\%, it may simply introduce more complexity to consider the delensing operation for mitigating baryonic biases, and one may prefer to simply use a forward model of the nonlinear lensing power as we considered in the previous two subsections.  We leave to future work the calculation of the baryonic feedback biases from delensed CMB power spectra.

\section{User-induced systematics: miscalculating of $C_\ell$}\label{sec:inaccuracy_bias}

The biases considered thus far are physical, in the sense that they arise from inaccuracy in the nonlinear lensing model due to our imperfect knowledge of non-perturbative baryonic physics.  However, similar biases can also arise due to purely numerical accuracy issues in theoretical calculations.  In this section, we show that it is imperative to use higher numerical accuracy than is currently standard in calculations of $C_\ell^\lensed$ when performing high-precision CMB data analysis.   In particular, the default accuracy parameters used in calculating $C_\ell^\lensed$ in the most widely-used Einstein-Boltzmann codes, CAMB and CLASS, are not adequate for SO- and CMB-S4-like analyses. Here, we compute the bias one would obtain from an analysis of SO or CMB-S4 primary CMB data when using the \textit{default} accuracy parameters of CAMB. 

For clarity, it will be useful in this section to show explicitly the code we use to calculate $C_\ell^\lensed$  with CAMB. The power spectra are calculated from an instance of the pyCAMB class \texttt{CAMBparams} as follows:
\begin{widetext}
\begin{verbatim}
pars = camb.CAMBparams()
pars.set_cosmology(H0=67.32, ombh2=0.022383, omch2=0.12011, mnu=0.06, omk=0, tau=0.0543)
pars.InitPower.set_params(As=2.1e-9, ns=0.96605, r=0)
pars.NonLinear = model.NonLinear_both
pars.NonLinearModel.set_params(`mead2016', HMCode_A_baryon = 3.13, HMCode_eta_baryon = 0.603)
pars.set_for_lmax(10000, lens_potential_accuracy=8, lens_margin=2050);
pars.set_accuracy(AccuracyBoost=2.0, lSampleBoost=2.0, lAccuracyBoost=2.0,
  DoLateRadTruncation=False);                                    
results = camb.get_results(pars)
powers = results.get_cmb_power_spectra(pars)
totCL = powers[`total'].
\end{verbatim}
\end{widetext}
The default values of the accuracy parameters are much lower, being in fact
\begin{verbatim}
pars.set_for_lmax(lmax,
 lens_potential_accuracy=0, lens_margin=150);
pars.set_accuracy(AccuracyBoost=1.0,
 lSampleBoost=1.0, lAccuracyBoost=1.0, 
 DoLateRadTruncation=True);                                    
\end{verbatim}
although the pyCAMB documentation\footnote{\url{https://camb.readthedocs.io/}} notes that \texttt{lens\_potential\_accuracy=1} is necessary for \emph{Planck}-level accuracy in the lensing potential calculation.

We define our $C_\ell^{\mathrm{fiducial}}$ by calculating the $C_\ell$ using these default parameters, except with \texttt{lens\_potential\_accuracy=1} and \texttt{lmax} = 5000 as opposed to 10000; the latter further degrades the accuracy by a significant amount, even for $\ell<5000$.  We use derivatives of this quantity when calculating the Fisher matrix in Eq.~\eqref{fisher_matrix}.

\begin{figure}[!t]
\includegraphics[width=\columnwidth]{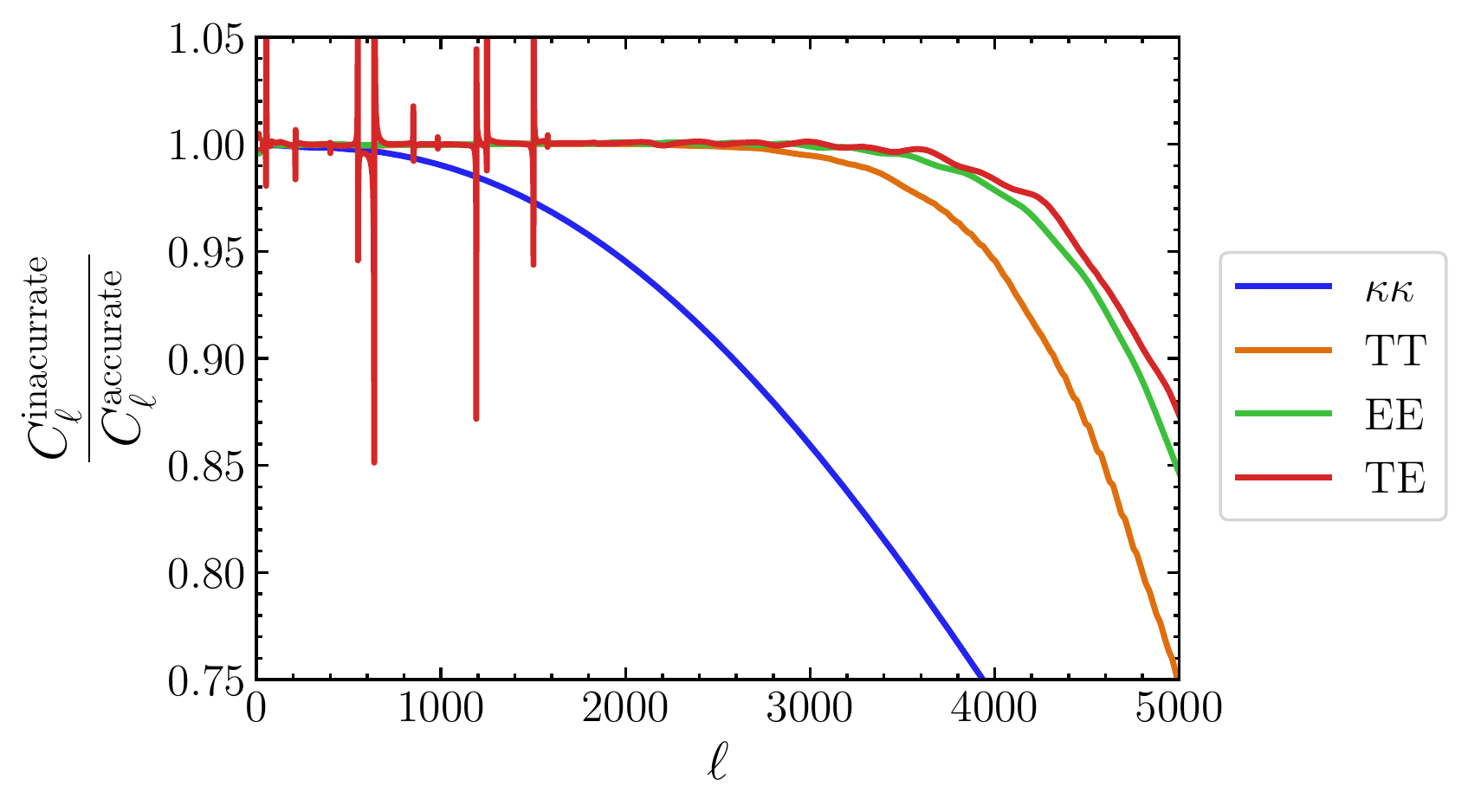}
\caption{The level of inaccuracy in the power spectra caused by using the default parameters in CAMB, as quantified by the ratio of the power spectra computed with default precision settings to those computed with high-precision settings.  The blue curve shows the CMB lensing convergence power spectrum, while the others show the lensed primary CMB power spectra, as labeled.}\label{fig:inaccurate_power}
\end{figure}

We show the inaccuracies in the lensing convergence power spectrum and the lensed power spectra, when run at these lower-precision settings, in Figure \ref{fig:inaccurate_power}.  By coincidence, these (purely numerical) effects are similar to those seen for the (physical) baryonic suppression in Fig.~\ref{fig:baryon_suppression_allsims}, but here they are larger in magnitude, reaching 25\% for the lensing convergence power spectrum at $L=4000$.

\begin{figure}[!t]
\includegraphics[width=\columnwidth]{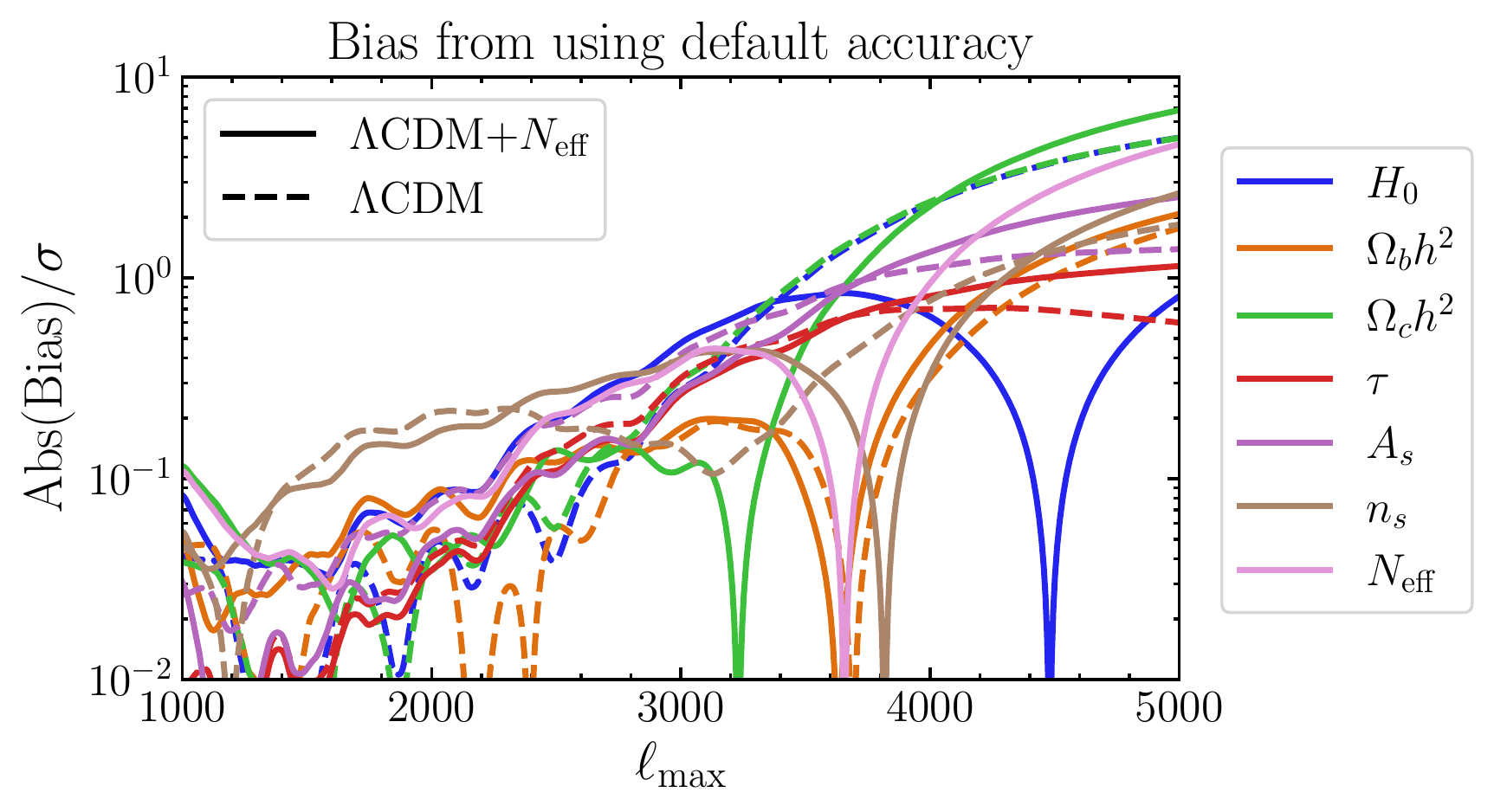}
\caption{The bias on each parameter, if the default accuracy CAMB settings were used to model the CMB power spectra in the analysis of CMB-S4 data (see Figure~\ref{fig:inaccurate_power}). The bias is shown as a fraction of the forecast 1$\sigma$ error.}\label{fig:inaccurate_biases}
\end{figure}

Using the formalism from Sec.~\ref{sec:fisher}, we calculate the biases induced in the inference of the $\Lambda$CDM or $\Lambda$CDM+$\Neff$ parameters if the analysis were to be done with the incorrect $C_\ell$ shown in Figure~\ref{fig:inaccurate_power}.  The biases are shown as a function of the maximum multipole included in the analysis $\ell_{\mathrm{max}}$ in Figure~\ref{fig:inaccurate_biases} ($\ell_{\rm max}$ is assumed to be the same for $TT$/$TE$/$EE$ here).  It is clear that the default accuracy settings are insufficient for these low-noise, high-resolution experiments, and their use would bias significantly any parameter inferences, with systematic errors as large as $\approx 5$ -- $6\sigma$ on $\Omega_c h^2$ and $H_0$, and $4\sigma$ on $\Neff$.  Fortunately, it is straightforward to remedy these biases by running the Einstein-Boltzmann codes with increased numerical accuracy settings.  For computational efficiency in MCMC analyses, one should determine the minimal accuracy settings needed to obtain sufficiently accurate predictions given the data under consideration (as done for CLASS, considering \emph{Planck} data, in Ref.~\cite{2011arXiv1104.2934L}).  We leave the detailed determination of the optimal settings, given computational constraints, to future analyses focused on each particular experiment.  Our results indicate that attention should be paid to this issue for SO and CMB-S4, and likely for the analysis of upcoming ACT and South Pole Telescope data as well, given the large biases seen in Figure~\ref{fig:inaccurate_biases}.

\section{Discussion and Conclusions}
\label{sec:discussion}

In this work, we have shown that inadequate modeling of baryonic feedback can lead to significant biases on cosmological parameters inferred from the primary CMB power spectra.  The biases enter through the gravitational lensing contribution in the damping tail of the CMB power spectra, which in turn is dependent on the matter power spectrum and hence susceptible to mismodeling of non-linear and baryonic feedback effects. As can be seen from Table \ref{tab:biases_SOS4}, for the number of light relativistic species $N_{\rm{eff}}$, the biases are as large as 0.38$\sigma$ (1.2$\sigma$) for SO (CMB-S4). For the Hubble constant $H_0$ in a fit to the $\Lambda$CDM model, they are as large as 0.96$\sigma$ (1.62$\sigma$) for SO (CMB-S4). These biases are estimated by assuming that the OWLS-AGN baryonic feedback model is the true model that describes the matter power spectrum in our universe, while the parameter inference is performed assuming no baryonic feedback prescription.   The OWLS-AGN model is a reasonable prescription to consider in this context, given the large spread of predictions from various subgrid prescriptions and AGN feedback strengths in modern hydrodynamical simulations (see, e.g.,~\cite{2019JCAP...03..020S,vanDaalen:2019pst}).

We have suggested multiple mitigation methods to avoid these uncertain late-universe effects in the primary CMB. Our first recommendation is to explicitly discard all data at $\ell>3000$ in the $TT$ spectra; we find that this choice reduces the biases on parameters considered here to be no more than 30\% of the statistical error bar (for both SO and CMB-S4). Alternatively, or in addition, we show that one can marginalize over a two-parameter model describing the effects of baryonic feedback. We find that this procedure reduces the biases on cosmological parameters by factors of $\mathcal{O}(100-1000)$.
For both of these mitigation strategies, the size of the statistical uncertainties on cosmological parameters increases, albeit not dramatically (generally $\lesssim$ 10\%, with a maximum increase of 21\% on $\Omega_c h^2$; the $\Neff$ error bar increases by 13\% for CMB-S4).  For the baryonic-parameter-marginalization approach, the increase could be mitigated by performing a joint analysis with $C_L^{\kappa\kappa}$.  It is conceivable that delensing will be useful as a data-driven solution --- that is, a solution without marginalization over a baryonic feedback model --- to avoid these biases. However, this will require significant delensing of the small-scale maps ($\ell>3000$); we leave to future work a quantitative study of the effectiveness of delensing with efficiencies expected for SO and CMB-S4 in combination with {\it Planck} CIB and galaxy surveys. Finally, we note that in the coming decade, cross-correlations between CMB experiments like SO and CMB-S4 and galaxy surveys like DESI and LSST will allow for percent-level measurements of the distribution of ionized electrons across a wide range of redshifts through the kinematic Sunyaev-Zel'dovich (kSZ) effect (e.g., \cite{1504.03339,1510.06442,Hill:2016dta,1603.03904,2016PhRvD..94l3526F,1607.02139,2009.05557,2009.05558,2101.08373,2101.08374,2021arXiv210201068K}).  In tandem, percent-level measurements of the ionized gas pressure across a wide range of halo masses and redshifts will be enabled by cross-correlations of these galaxy surveys with thermal Sunyaev-Zel'dovich (tSZ) maps~(e.g.,~\cite{Ade:2012nia,2015ApJ...808..151G,2016ApJ...819..128S,2017MNRAS.467.2315V,2018PhRvD..97h3501H,2018ApJ...865..109S,2019PhRvD.100f3519P,2020MNRAS.491.2318T,2020PhRvD.101d3525P,2020MNRAS.491.5464K}), enabling joint constraints on all thermodynamic quantities describing the ionized gas in and around galaxies~\cite{2017JCAP...11..040B}.  A wide variety of other probes will also be crucial in this endeavor, including X-ray measurements, fast radio burst dispersion measures, absorption line measurements, intensity-mapping measurements, and more.  These measurements will significantly reduce uncertainties on baryonic feedback models, thus motivating joint analyses of CMB power spectra and the kSZ and tSZ effects, as well as folding in external information from the full complement of baryonic probes. 

Given the non-negligible size of the baryonic-feedback biases for upcoming experiments, it is possible that ongoing experiments like ACT and the South Pole Telescope (SPT) that probe the CMB damping tail could also be mildly impacted by these effects.  While recent SPT analyses have either used $\ell_{\rm max}=3000$ \cite{2011ApJ...743...28K} or excluded $TT$ data altogether \cite{2018ApJ...852...97H, 2021arXiv210101684D}, the recent ACT DR4 analysis \cite{2020JCAP...12..047A,Choi:2020ccd} used $TT$ data out to $\ell=4325$. However, we expect any biases in analyses to date to be well below the $0.2\sigma$ level given that (a) the uncertainties on $H_0$ are more than five times larger than the forecast for SO (with a bias from baryonic feedback of $\Delta H_0=0.96\sigma$, c.f. Table~\ref{tab:biases_SOS4}) and (b) the absolute bias should also be lower than found here since the instrument noise level is larger in comparison with the gravitational lensing contribution to the power spectra.  We also emphasize that these biases are irrelevant for \emph{Planck} CMB power spectra, which do not probe multipoles $\ell \gtrsim 3000$ where the baryonic effects become important.  These considerations do, however, highlight that mitigation strategies should be adopted for upcoming analyses from Advanced ACT~\cite{Henderson2016} and SPT-3G \cite{spt3g,spt3g2,spt3g3}.

We have also investigated biases that would arise if (default) low-accuracy settings in Einstein-Boltzmann codes are used to calculate the CMB power spectra.  These biases are highly significant (up to $6\sigma$ on some cosmological parameters), highlighting the need for care when using these codes, as well as the need for a systematic study of the accuracy parameters required for upcoming experiments.  Our results also motivate a new, detailed comparison between CAMB and CLASS in preparation for high-precision, high-resolution CMB surveys like SO and CMB-S4, as we have shown that the data will be sensitive to effects that had previously escaped attention.

Our work has only considered the $\Lambda$CDM parameters and the $N_{\rm eff}$ parameter, but similar considerations may apply to other parameters that affect the damping tail of the CMB. Inference of the sum of the neutrino masses may be of concern, but we note that the dominant constraint on this parameter comes from a more direct reconstruction of the gravitational lensing signal through the four-point function of the CMB, where the contribution from small scales in the matter power spectrum is easier to control (see \cite{2020arXiv201106582M} for a detailed study). Inference of blackbody secondary anisotropy parameters like the amplitude of the kSZ power spectrum (both late-time as well as from the reionization epoch) could also be affected, as these parameters can be degenerate with the primary cosmological parameters (which can be biased, as we have shown). For the same reason, free parameters in the blackbody components in the foreground model (e.g., the kSZ power spectrum amplitude) could also mitigate the baryonic feedback biases considered in this work to some extent, by absorbing their effects (at the cost of a biased inference of the kSZ amplitude). We leave investigation of these issues to future work.

Several decades of cosmological inference from the primary CMB have benefited from the simplicity of the linear physics responsible for the observed temperature and polarization anisotropies. This will change in the coming decade with the next generation of CMB surveys. While these surveys extract new cosmological and astrophysical information from late-time effects, the primary CMB signal also becomes increasingly sensitive to uncertain astrophysical phenomena. A careful consideration of mitigation strategies, including delensing and/or joint analyses with the kSZ and tSZ effects, will therefore be of great importance in the coming decade of CMB surveys.

\noindent

\begin{acknowledgements}
We thank Anthony Challinor, Antony Lewis, Neelima Sehgal, Blake Sherwin, and Alexander van Engelen for useful exchanges. We thank Eegene Chung and Simon Foreman for sharing their code to compute $P_m^{\mathrm{bary}}(k,z)$ for various simulations. JCH thanks the Simons Foundation for support. FMcC acknowledges support from the Vanier Canada Graduate Scholarships program. Research at Perimeter Institute is supported in part by the Government of Canada through
the Department of Innovation, Science and Industry Canada and by the Province of Ontario through the Ministry of Colleges and Universities.  This is not an official Simons Observatory Collaboration paper.
\end{acknowledgements}

\appendix

\section{Bias results for $\ell_{\rm max}^{TT} = 3000$ cutoff and baryonic parameter marginalization approaches}
\label{app:biases}

In this Appendix we present the numerical values of the biases on each parameter (for the OWLS-AGN simulation, with either the SO or CMB-S4 configuration) after applying the mitigation methods; the $\ell_{\mathrm{max}}^{TT}=3000$ results are shown in Table \ref{tab:biases_SOS4_lmax3000} and the marginalization results are shown in Table \ref{tab:biases_SOS4_aeta}.

\begin{table}[h!]
\begin{tabular}{|c|c|c|c|c|}\hline
&\multicolumn{2}{c|}{SO ($\ell_{\mathrm{max}}^{TT}=3000$)}&\multicolumn{2}{c|}{CMB-S4 ( $\ell_{\mathrm{max}}^{TT}=3000$)}\\\cline{2-5}
&$\Lambda$CDM&$\Lambda$CDM$+\Neff$&$\Lambda$CDM&$\Lambda$CDM$+\Neff$\\\hline\hline
$H_0$   &   0.098   &   0.25   &   0.18   &   0.23\\\hline
$\Omega_bh^2$   &   0.13   &   0.085   &   0.15   &   0.065\\\hline
$\Omega_ch^2$   &   0.12   &   0.12   &   0.20   &   0.05\\\hline
$	\tau$   &   0.15   &   0.11   &   0.19   &   0.15\\\hline
$A_s$   &   0.18   &   0.076   &   0.23   &   0.15\\\hline
$n_s$   &   0.016   &   0.19   &   0.0084   &   0.18\\\hline
$N_{\mathrm{eff}}$   &      &   0.24   &      &   0.23\\\hline

\end{tabular}
\caption{Fractional biases from the OWLS-AGN model (in units of the forecast $1\sigma$ statistical error bar) on each of the parameters in the various setups, when an $\ell_{\mathrm{max}}^{TT}=3000$ cutoff is imposed (to be compared with Table \ref{tab:biases_SOS4}.)}\label{tab:biases_SOS4_lmax3000}
\end{table}

\begin{table}[h!]
\begin{tabular}{|c|c|c|c|c|}\hline
&\multicolumn{2}{c|}{SO }&\multicolumn{2}{c|}{CMB-S4}\\\cline{2-5}
&$\Lambda$CDM&$\Lambda$CDM$+\Neff$&$\Lambda$CDM&$\Lambda$CDM$+\Neff$\\\hline\hline
$H_0$   &   0.0028   &   0.0022   &   0.0030   &   0.0026\\\hline
$\Omega_bh^2$   &   0.00048   &   0.00098   &   0.00030   &   0.00060\\\hline
$\Omega_ch^2$   &   0.0028   &   0.00091   &   0.0031   &   0.0012\\\hline
$	\tau$   &   0.0026   &   0.0025   &   0.0038   &   0.0036\\\hline
$A_s$   &   0.0035   &   0.0031   &   0.0051   &   0.0046\\\hline
$n_s$   &   0.0014   &   0.0015   &   0.0016   &   0.0018\\\hline
$N_{\mathrm{eff}}$   &      &   0.00091   &      &   0.0011\\\hline

\end{tabular}
\caption{Fractional biases from the OWLS-AGN model (in units of the forecast $1\sigma$ statistical error bar) on each of the parameters in the various setups, when the baryonic parameters $A,\eta_0$ are marginalized over (to be compared with Table~\ref{tab:biases_SOS4}).}\label{tab:biases_SOS4_aeta}
\end{table}

\begin{table*}[h!]
    \centering
    \begin{tabular}{|c||c|c|c||c|c|c|}\hline
    
    &\multicolumn{3}{c||}{$1\sigma$ error [\%]: SO}&\multicolumn{3}{c|}{$1\sigma$ error [\%]: CMB-S4}\\\hline
    &$\ell_{\mathrm{max}}^{TT}=5000$&$\ell_{\mathrm{max}}^{TT}=3000$&$\ell_{\mathrm{max}}^{TT}=5000+A+\eta$&$\ell_{\mathrm{max}}^{TT}=5000$&$\ell_{\mathrm{max}}^{TT}=3000$&$\ell_{\mathrm{max}}^{TT}=5000+A+\eta$\\\hline\hline
        $H_0$   &   0.40   &   0.43   &   0.44   &   0.34   &   0.37   &   0.38\\\hline
$\Omega_bh^2$   &   0.22   &   0.23   &   0.23   &   0.15   &   0.15   &   0.16\\\hline
$\Omega_ch^2$   &   0.58   &   0.63   &   0.64   &   0.50   &   0.55   &   0.56\\\hline
$\tau$   &   10   &   10   &   10   &   9.3   &   9.3   &   9.5\\\hline
$A_s$   &   0.95   &   0.97   &   1.0   &   0.86   &   0.87   &   0.90\\\hline
$n_s$   &   0.26   &   0.27   &   0.27   &   0.23   &   0.23   &   0.23\\\hline
    \end{tabular}
    \caption{Forecast constraints (as percentages of the fiducial parameter values) for the $\Lambda$CDM parameters, for the various mitigation methods for SO and CMB-S4.}
    \label{tab:LCDM_forecast_full}
\end{table*}

\begin{table*}[h!]
    \centering
    \begin{tabular}{|c||c|c|c||c|c|c|}\hline
    
    &\multicolumn{3}{c||}{$1\sigma$ error [\%]: SO}&\multicolumn{3}{c|}{$1\sigma$ error [\%]: CMB-S4}\\\hline
    &$\ell_{\mathrm{max}}^{TT}=5000$&$\ell_{\mathrm{max}}^{TT}=3000$&$\ell_{\mathrm{max}}^{TT}=5000+A+\eta$&$\ell_{\mathrm{max}}^{TT}=5000$&$\ell_{\mathrm{max}}^{TT}=3000$&$\ell_{\mathrm{max}}^{TT}=5000+A+\eta$\\\hline\hline
        $H_0$   &   0.74   &   0.75   &   0.82   &   0.58   &   0.60   &   0.63\\\hline
$\Omega_bh^2$   &   0.33   &   0.33   &   0.33   &   0.22   &   0.23   &   0.23\\\hline
$\Omega_ch^2$   &   0.95   &   1.1   &   1.1   &   0.72   &   0.87   &   0.87\\\hline
$\tau$   &   10   &   10   &   10  &   9.3   &   9.4   &   9.5\\\hline
$A_s$   &   1.0   &   1.0   &   1.1   &   0.89   &   0.91   &   0.92\\\hline
$n_s$   &   0.47   &   0.48   &   0.50   &   0.39   &   0.40   &   0.41\\\hline
$N_{\mathrm{eff}}$   &   2.0   &   2.2   &   2.3   &   1.5   &   1.7   &   1.7\\\hline
    \end{tabular}
    \caption{Forecast constraints (as percentages of the fiducial parameter values) for the $\Lambda$CDM$+\Neff$ model parameters, for the various mitigation methods for SO and CMB-S4.}
    \label{tab:LCDMNeff_forecast_full}
\end{table*}

\clearpage

\bibliography{references}

\end{document}